\documentclass{optica-article}

\journal{opticajournal} 

\articletype{Review Article}

\usepackage{lineno}

\begin{document}

\title{Metasurfaces in Adaptive Optics: A New Opportunity in Optical Wavefront Sensing}

\author{Rundong Fan,\authormark{1,2,3}, Zichao Wang,\authormark{1,2,3}, Pei Li,\authormark{1,2,3} and Lei Huang,\authormark{1,2,3*}}

\address{\authormark{1}Key Laboratory of Photonic Control Technology (Tsinghua University), Ministry of Education, Beijing 100084, China\\
\authormark{2}State Key Laboratory of Precision Space-time Information Sensing Technology, Beijing 100084, China\\
\authormark{3}Department of Precision Instrument, Tsinghua University, Beijing 100084, China}

\email{\authormark{*}hl@tsinghua.edu.cn} 


\begin{abstract*} 
Over the past fifty years, wavefront sensing technology has continuously evolved from basic techniques to high-precision systems, serving as a core methodology in adaptive optics (AO). Beyond traditional wavefront retrieval methods based on spot displacement, direct phase retrieval techniques with greater accuracy have emerged, jointly driving advancements in wavefront sensing precision. This evolution is fueled by increasing demands for accuracy, which have prompted iterative upgrades in system architectures and algorithms. Recently, breakthroughs in metasurface technology have opened new possibilities for wavefront sensing. By utilizing subwavelength microstructures, metasurfaces enable multi-dimensional control over the phase, amplitude, and polarization of light fields. Their high degree of design flexibility presents transformative opportunities for advancing wavefront sensing capabilities. This review examines the fundamental principles of wavefront sensing and the development of key enabling devices, highlighting how metasurface technology is reshaping traditional paradigms. We discuss recent research progress and emerging innovations, aiming to establish a theoretical framework for next-generation wavefront sensing technologies. Ultimately, we hope this review provides technical insights for applications in astronomical observation, biological microscopy, laser engineering, and beyond. 

\end{abstract*}

\section{Introduction}
 Astronomers have been observing celestial phenomena through ground-based telescopes for centuries. However, the refractive distortion of light as it passes through Earth's atmosphere results in blurred images. The advent of adaptive optics (AO) has mitigated this issue. An AO system is defined as an optical system designed for the real-time measurement and correction of optical aberrations, combining photoelectric and computational technologies \cite{hampson2021adaptive,hampson2008adaptive}. The concept of an adaptive optics system was first proposed by Babcock in 1953 and applied to ground-based telescopes to compensate for the effects of atmospheric turbulence on image quality, marking a significant advancement in astronomical observation \cite{babcock1953possibility,babcock1990adaptive}. In 1993, Beckers employed the AO system to achieve diffraction-limited imaging, significantly improving angular resolution and sensitivity \cite{beckers1993adaptive}. This breakthrough surpassed the arc-second resolution limit imposed by Earth's atmosphere, while Beckers also provided a comprehensive overview of the history of adaptive optics systems. Owing to their exceptional performance in astronomy, adaptive optics systems have been progressively adopted across various fields, including biomedical imaging \cite{ji2012characterization,liu2019direct,miller2020cellular,hofer2011wavefront,kong2024adaptive,yao2023construction, booth2007adaptive}, laser shaping \cite{salter2019adaptive,bechet2014beam,arain2007adaptive,yang2007adaptive,nie2021adaptive,he2023vectorial}, laser communication \cite{zhang2022application,horst2023tbit,wang2022orbital,chen2019experimental,li2017adaptive}, and laser weapons \cite{extance2015laser,scheers2023numerical}.

 The core principle of an adaptive optics system is to measure wavefront distortion in real time and use at least one deformable optical element to correct optical aberrations. Typically, an adaptive optics system comprises three key components: 1) At least one wavefront sensor to detect optical aberrations caused by disturbances (such as atmospheric turbulence, cellular structures, etc.); 2) At least one wavefront corrector (like a deformable mirror or spatial light modulator (SLM)) to rectify these aberrations; and 3) A real-time control system that generates commands for the wavefront corrector based on the optical aberrations measured by the wavefront sensors, allowing the system to track changes in the distorted wavefront.
As the front end of the adaptive optics system, the wavefront sensor is a key factor determining the system's correction capability. Consequently, extensive research has been conducted in this field to optimize the performance of wavefront sensors. Traditional wavefront sensors are generally classified into three categories: curvature wavefront sensors \cite{zepp2013holographic,paterson2000hybrid,hickson1994single,cagigal2015x,liu2025high,diaz2006curvature,letchev2023assessing,goloborodko2023wavefront,zhong2022hybrid}, pyramid wavefront sensors \cite{frazin2018efficient,guthery2021pyramid,agapito2023non,guzman2024deep,bond2020adaptive,bertrou2022confusion,chambouleyron2020pyramid,lozi2019visible,akondi2013digital,ragazzoni2002pyramid}, and Shack-Hartmann wavefront sensors \cite{meimon2014sensing,aftab2018adaptive,basavaraju2014myopic,deng2025measurement,he2024accuracy,li2021piston,zhang2024automatic,yang2022method,xiya2023research,hartlieb2024large,wu2023study,deng2025sequential,zheng2021detecting}. With advances in computational technology, phase retrieval techniques for wavefront sensing at the exit pupil of optical systems have been increasingly adopted, particularly in the field of phase imaging \cite{wang2021deep,wang2015direct,yi2021angle,berto2017wavefront,wu2019wish,norris2020all,guo2024direct}. With improvements in geometric optics and algorithms, the dynamic range and accuracy of wavefront sensors have seen significant enhancements. However, the ability of geometric optical elements to regulate light remains limited, and they are still unable to effectively overcome the dynamic range and detection accuracy constraints imposed by the physical properties of traditional optical elements (such as radius of curvature and microstructure). As a result, modulation of light on the nanoscale for wavefront detection has become a key focus in the development of modern optics and nanophotonics.

A metasurface is a two-dimensional sub-wavelength material composed of periodic, quasi-periodic, or randomly arranged artificial nanostructures resonantly coupled with incident electromagnetic waves. This enables the metasurface to precisely control the amplitude, phase, polarization, and dispersion of the light field at the nanoscale, facilitating the realization of electromagnetic effects that are difficult to achieve with conventional optical components \cite{li2025orthogonal,kamali2018review,zhang2020polarization,balthasar2017metasurface,dorrah2021metasurface,yu2014flat,huang2013three,zhao2020recent,zheng2015metasurface,deng2020malus,yue2022versatile,wang2021resonant,hao2021full,high2015visible}. Metasurfaces can be manufactured using modern techniques such as nanoimprint lithography \cite{park2025tape,einck2021scalable,chen2015large,makarov2017multifold,choi2023realization}, laser direct writing \cite{chen2022large,hu2021high,nivas2025femtosecond,sun2025generating,zhang2025recent}, and electron beam etching \cite{sin2023high,baracu2021silicon,cao2022tuning,briere2019etching,baracu2021metasurface}. The significant advantage of metasurfaces is their ability to achieve high-resolution phase modulation across the full 2$\pi$ range at the optically thin interface, while also providing flexible modulation of both the wavefront of the optical field \cite{he2020metasurfaces,wei2020optical,zhao2018continuously,huang2024microcavity,wu2024ultrathin,barton2021wavefront}. The wavefront modulation of metasurfaces can be primarily classified into three categories: 1) Nano-antennas with different structures exhibit distinct resonances, and precise wavefront control can be achieved by adjusting the shape and geometric parameters of each nano-antenna \cite{bomzon2001pancharatnam,mehmood2016visible,khorasaninejad2015broadband,khorasaninejad2016polarization,aieta2015multiwavelength,kang2012wave}. 2) Wavefront modulation is achieved by tailoring the electromagnetic polarizability within the nanostructures, where each nanostructure block acts as a secondary wave source. This modulation principle follows the Huygens principle, classifying this type of metasurface as a Huygens metasurface \cite{chen2018huygens,wong2018perfect,leitis2020all,fan2018phototunable,gigli2021fundamental,eleftheriades2022prospects,jia2015independent,howes2018dynamic,londono2018broadband}. 3) Based on the Pancharatnam-Berry (PB) phase \cite{pancharatnam1956generalized,berry1987adiabatic}, incident light with the same polarization state experiences different phase shifts when it transitions to the same final polarization state via different paths in the Poincaré sphere \cite{wang2023metasurface,ren2015generalized,yi2015hybrid}. The phase of polarized light can be controlled by adjusting the rotation angle of the microstructure. Such metasurfaces typically require altering the geometry of the nanoantenna to achieve a half-wave plate design, commonly used to modulate circularly polarized light. This is in the field called the PB phase, or geometric phase \cite{lin2018polarization,ding2015ultrathin,tymchenko2015gradient,xie2021generalized,mcdonnell2021functional,choudhury2017pancharatnam,wang2022nonlinear,xu2022mechanically}. 

The development of metasurfaces offers a new perspective on the design of various optical systems. By integrating metasurfaces with nanostructures, many traditional phase modulation and wavefront shaping optical systems have achieved breakthroughs in functionality, performance, and compactness, leading to the creation of numerous new optical systems,such as spectrometers \cite{cai2024compact,ji2025compact,wang2023compact,faraji2018compact,tang2024metasurface,wen2024metasurface}, telescopes \cite{liu2020optical,zhang2022high,wang2025portable}, holographic displays \cite{gopakumar2024full,choi2025roll,song2021large,wang2024holographic,xiong20253d,fan2024integral,fan2024dual,bouchal2019high}, eyepieces \cite{wirth2025wide,wang2021metalens,li2022ultracompact,lee2018metasurface,ko2024metasurface} and so on. Wavefront sensing is the inverse problem of wavefront modulation. Metasurfaces composed of nanoantennas offer new opportunities for the development of high-precision wavefront sensors. This review first overview the regulatory principles of metasurface-based micro–nano optics, followed by a discussion of the operating mechanisms and current research progress in metasurface-enabled wavefront sensing technologies. Finally, we outline potential future directions for the development of metasurface-based wavefront sensing.

\section{Light Field Manipulation via Subwavelength Structured}

The metasurface is composed of an array of microstructures with dimensions and spacings smaller than the operating wavelength-commonly referred to as the effective wavelength of the system. A key feature of such structures is their ability to impart spatially varying phase distributions by tailoring individual elements' size, shape, and arrangement. Metasurfaces are often classified as planar optical devices due to their subwavelength thickness. The convergence of metamaterials and micro–nano fabrication technologies has enabled metasurfaces to modulate multiple degrees of freedom of the light field, including phase and polarization.

In the domain of light-field manipulation, current research predominantly focuses on dielectric metasurfaces. Readers seeking comprehensive overviews of the development and state-of-the-art in dielectric metasurfaces may refer to the reviews by Genevet et al. (2017) \cite{genevet2017recent}, Kamali et al. (2018) \cite{kamali2018review}, and Qiao et al. (2018) \cite{qiao2018recent}. With continued advances in the field, the functionality of metasurfaces has expanded beyond conventional control over phase, dispersion, and polarization, moving into emerging regimes such as nonlinear optics. Notable examples include second-harmonic generation\cite{li2017nonlinear,wang2017metasurface,ma2021nonlinear}, quasi-bound states in the continuum \cite{liu2018extreme}, and other phenomena that extend the capabilities of light-matter interaction.

While the focus of this review is on metasurface-enabled wavefront sensing-primarily involving the manipulation of phase and polarization, it is worth noting that a growing body of literature also explores metasurface-assisted sensing in broader contexts.  For developments in nonlinear metasurface optics, readers are directed to the review by Chen et al. \cite{chen2018phase}.

The physical mechanism by which optical elements manipulate light typically relies on phase accumulation to bend and shape wavefronts. Similarly, metasurfaces composed of subwavelength structural materials achieve light focusing and beam shaping through engineered phase accumulation. The discontinuous arrangement of nanostructures in metasurfaces introduces abrupt phase shifts, thereby providing new degrees of freedom for light-field modulation. In this section, we review the generalized Snell’s law, a principle commonly employed in wavefront sensing. A detailed explanation and derivation can be found in the research work by Yu et al. (2011) \cite{yu2011light}. 

The propagation of light adheres to Fermat’s principle, which states that the path taken by light between two points, A and B, corresponds to the one with the shortest optical path length. In its phase-stationary form, Fermat’s principle can be expressed as $\int_{A}^{B} d \varphi(\vec{r})$, that is, the derivative of the accumulated phase along the actual optical path with respect to a small variation in the path is 0. The introduction of a sudden phase shift, a phase discontinuity at the interface between two media allows a reinterpretation of the laws of reflection and refraction through the lens of Fermat’s principle. Consider a plane wave incident at an angle. If two paths lie infinitesimally close to the actual optical path (Fig. 1), the phase difference between them vanishes, which can be expressed as follows:
\begin{eqnarray}
\int_{A}^{B} d \varphi(\vec{r})\left[k_{\mathrm{o}} n_{\mathrm{i}} \sin \left(\theta_{\mathrm{i}}\right) d x+(\varphi+d \varphi)\right]-\left[k_{\mathrm{o}} n_{\mathrm{t}} \sin \left(\theta_{\mathrm{t}}\right) d x+\varphi\right]=0,
\end{eqnarray}
where $\theta_t$ is the angle of refraction,$\varphi$ and $\varphi+d\varphi$ represent the phase discontinuities at the respective points where the two paths intersect the interface. $dx$ denotes the distance between the intersection points along the interface; $n_i$ and $n_t$ are the refractive indices of the incident and transmission media, respectively; and $k_o$ = 2$\pi$/$\lambda_o$, where $\lambda_o$ is the wavelength in vacuum. When the phase gradient along the interface is engineered to be constant, the equation yields the generalized Snell’s law of refraction. It is expressed as follows:
\begin{equation}
\sin \left(\theta_{\mathrm{t}}\right) n_{\mathrm{t}}-\sin \left(\theta_{\mathrm{i}}\right) n_{\mathrm{i}}=\frac{\lambda_{\mathrm{o}}}{2 \pi} \frac{d \varphi}{d x},
\end{equation}
where  $d\varphi$⁄$dx$ is represented by a phase variation at position $x$. According to the generalized Snell’s law, refraction at arbitrary angles can be achieved by introducing a discontinuous phase shift, $d\varphi$⁄$dx$, at the optical interface. Moreover, due to the non-zero phase gradient in this modified law, incident angles of opposite sign can yield different refraction angles. When light propagates from a medium of higher refractive index to one of lower refractive index, two critical angles for total internal reflection may arise, expressed as follows:
\begin{equation}
\theta_{\mathrm{c}}=\arcsin \left( \pm \frac{n_{\mathrm{t}}}{n_{\mathrm{i}}}-\frac{\lambda_{\mathrm{o}}}{2 \pi n_{\mathrm{i}}} \frac{d \varphi}{d x}\right).
\end{equation}

Furthermore, for reflective optical systems, the generalized Snell’s law is as follows:
\begin{eqnarray}
\sin \left(\theta_{\mathrm{r}}\right)-\sin \left(\theta_{\mathrm{i}}\right) & = & \frac{\lambda_{\mathrm{o}}}{2 \pi n_{1}} \frac{d \varphi}{d x},
\end{eqnarray}
where $\theta_i$ and $\theta_r$ represent the incidence and reflection angles, respectively. Unlike specular reflection, the relationship between the incident and reflected angles governed by the generalized Snell’s law is nonlinear. A critical state exists between the incident and reflects angles, which can be described as follows:
\begin{eqnarray}
\theta_{\mathrm{c}}^{\prime} & = & \arcsin \left(1-\frac{\lambda_{\mathrm{o}}}{2 \pi n_{\mathrm{i}}}\left|\frac{d \varphi}{d x}\right|\right) .
\end{eqnarray}

When the incidence angle exceeds the critical angle, the intensity of the reflected light diminishes significantly.

Metasurfaces rely on nanoantennas to introduce abrupt phase shifts at optical interfaces. However, the derivation of the generalized Snell's law assumes a continuous phase profile, that is, $d$$\varphi$⁄$d x=0$, such that all incident energy is directed into the anomalous reflection and refraction channels. In practice, the discrete arrangement of nanoantennas results in residual conventional reflection and refraction, which can degrade optical performance. By carefully engineering the size, shape, and spacing of the nanoantennas, the impact of these unwanted diffraction orders can be significantly suppressed.

The design of metasurface optics mainly involves selecting appropriate antenna geometries and spatial arrangements to achieve complete 2$\pi$ phase coverage while minimizing the detrimental effects of conventional refraction. A widely adopted approach is to construct a nanoantenna library through parameter sweeps, followed by efficient spatial arrangement using numerical optimization techniques. Recently, with advances in computational power and artificial intelligence, end-to-end design frameworks for metasurfaces have emerged. A comprehensive overview of metasurface design methodologies can be found in recent reviews, such as that published in 2022 \cite{so2023revisiting}. Through advances in nanoantenna engineering, metasurfaces can not only achieve anomalous light deflection, but also modulate the polarization state of the light field. In 2012, Chen and colleagues from the University of Birmingham demonstrated full 0–2$\pi$ phase control of circularly polarized light using rod-shaped nanometallic antennas with varying rotation angles. By arranging these antennas in a spatially ordered pattern at subwavelength scales, they successfully achieved spatial focusing of visible light fields \cite{chen2012dual}. This work revealed that a phase gradient from 0 to 2$\pi$ can be introduced in the polarization-converted output of circularly polarized light by sequentially rotating the nanoantennas from 0 to $\pi$ radians in a plane perpendicular to the incident wave vector, around their symmetry axes aligned along the propagation direction. The subwavelength-scale phase gradient generated through this structural rotation effectively redirects the wave vector of the outgoing circularly polarized light field.
\begin{figure}[ht]
\centering\includegraphics[width=8cm]{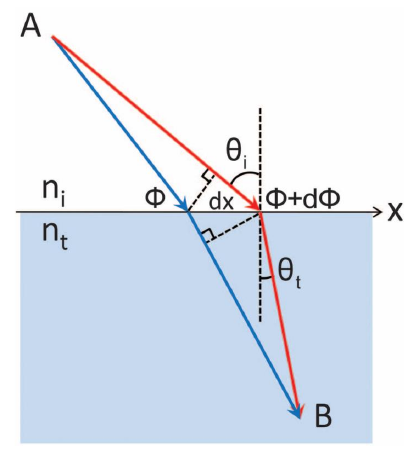}
\caption{A schematic derivation of the generalized Snell' s law is shown (Ref. \cite{yu2011light}, Fig. 1), in which the interface between two media is engineered to introduce a sudden phase discontinuity in the optical path, dependent on the position along the interface.$\varphi$ and $\varphi+d\varphi$ represent the phase differences at the locations where the two paths (blue and red) intersect the boundary.}
\end{figure}

\section{Wavefront Sensing Paradigms: From Classical Techniques to Metasurface-Driven Innovations}
The advancement of metasurface-based micro- and nano-optics offers unprecedented degrees of freedom for light field modulation and encoding. By leveraging a single or a limited number of optical elements, a wide range of optical functionalities can now be realized. This multifunctional integration provides novel avenues for wavefront sensing. The impact of metasurfaces on wavefront sensing is primarily reflected in three key areas: (1) replacing conventional optical components used for light field encoding with metasurfaces, enabling high-precision phase modulation and significantly enhancing both the measurement accuracy and dynamic range of wavefront sensors; (2) integrating multiple optical functions into a single metasurface, thereby simplifying complex optical architectures and paving the way for miniaturized wavefront sensing systems; and (3) harnessing the unique physical effects induced by subwavelength meta-atoms in metasurfaces, allowing the development of entirely new wavefront sensing schemes that transcend the performance limits of traditional approaches. To provide a comprehensive perspective on the latest developments in this rapidly evolving field, this section first categorizes existing wavefront sensing methodologies and summarizes recent advances. It then presents a review of metasurface-enabled approaches across each category. The detailed classification of wavefront sensing schemes is illustrated in Fig. 2.
\begin{figure}[ht]
\centering\includegraphics[width=12cm]{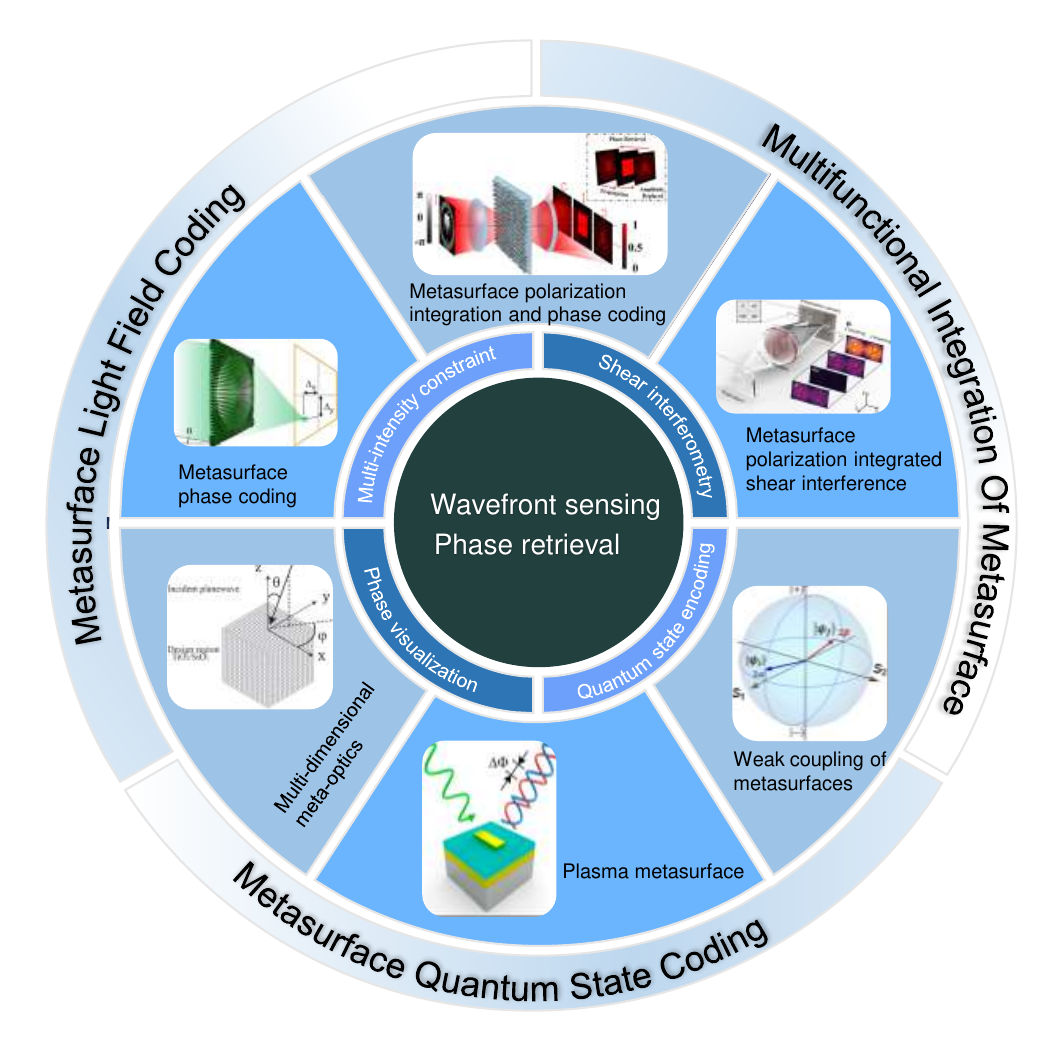}
\caption{Schematic classification of wavefront sensing schemes, encompassing Shack–Hartmann sensors, wavefront reconstruction via multi-intensity constraint methods, shear interferometry-based approaches, non-interferometric phase visualization techniques, and advanced schemes enabled by metasurfaces, including multi-dimensional control and quantum state encoding.}
\end{figure}

\subsection{Shack–Hartmann Wavefront Sensing: From Microlens Arrays to Metasurface-Based Light Field Modulation}

The Shack–Hartmann wavefront sensor is a device that divides a distorted wavefront into several sub-wavefronts using an array of optical elements and reconstructs the complete wavefront by analyzing the wavelet parameters. A prominent early application of this approach is the Hartmann sensor, initially developed by Roland Shack \cite{platt2001history}. In the earliest Hartmann wavefront sensing experiments, large plates or cardboard panels perforated with holes were placed over the aperture of large telescopes. These plates were positioned at known distances on either side of the focal plane, and each hole produced its own blurred image. By capturing two images at different positions and measuring the centroid of each spot, the propagation of light rays through the focal region could be traced. These ray traces were then used to assess the optical quality of large telescopes. The system consists of a working optical element, a pinhole array diaphragm, and two symmetrically distributed sensor elements. When an ideal plane wave is incident, the relative positions of the diffraction spots on the sensor elements remain fixed. However, in the presence of wavefront distortion, the positions of the diffraction spots shift. By measuring changes in the center of mass and the spacing of the diffraction spots for each aperture, the wavefront curvature can be quantitatively reconstructed.

However, due to the inherent structural limitations of the diaphragm array and the separate sensor, early Hartmann sensors suffered a loss of more than 60$\%$ light energy, with a typical aperture efficiency of only 38$\%$–42$\%$. This substantial loss severely restricted spot positioning accuracy. In 2001, Shack introduced revolutionary improvements, including: 1) Replacing the traditional mechanical diaphragm with a microlens array, integrating wavefront segmentation and focusing functions; 2) Optimizing the detector layout by positioning it at the focal plane. The resulting structural innovation significantly enhanced light energy utilization, enabling the detection of weak turbulent wavefronts.

The working principle of the Shack-Hartmann wavefront sensor (SHWFS)  can be divided into three key physical processes: 1) Wavefront segmentation – The incident wavefront is discretized into an $N$×$N$ sub aperture wavefront grid by the microlens array; 2) Focal plane imaging – Each sub-aperture wavefront is focused onto the CCD sensor by the microlenses, forming a spot array; 3) Distortion detection – When the sub-aperture wavefront deviates from the reference plane wave by an inclination angle $\theta_1$, the focal plane spots undergo a displacement $\Delta x$. The quantitative relationship between the wavefront slope and the spot displacement can be expressed as:
\begin{eqnarray}
\left\{\begin{array}{l}
g_{x}=\frac{\partial \phi(x, y)}{\partial x}=\frac{\Delta x}{f} \\
g_{y}=\frac{\partial \phi(x, y)}{\partial y}=\frac{\Delta y}{f}
\end{array}\right.
\end{eqnarray}
where $\phi$($x$,$y$) represents the phase distribution of the incident wavefront, while [$\Delta x$, $\Delta y$] denotes the relative displacement of the microlens focal spot in the $x$ and $y$ directions. The terms $g_x$ and $g_y$ correspond to the local slope components of the sub-aperture wavefront in the x and y directions, respectively. The wavefront reconstruction algorithm processes the slope measurements from all sub-apertures ($g_x$, $g_y$) and reconstructs the complete wavefront phase $\phi$($x$,$y$) using either the least-squares method or the modal reconstruction approach.

SHWFS are calibrated using plane wave incidence prior to operation to eliminate static wavefront aberrations introduced during fabrication and assembly. These sensors offer several advantages, including a simple structure, fast measurement speed, compatibility with various light sources, and high measurement accuracy within the dynamic range. However, their performance is constrained by the size of the microlens array, as the wavefront cannot be effectively detected when the focal spot displacement exceeds the sub-aperture region.  Additionally, ambient noise can interfere with slope calculations, leading to phase reconstruction errors.

Optimizing the calibration algorithm represents a key pathway for enhancing sensor performance. The evolution and key breakthroughs of SHWFS technology, driven by algorithmic optimization to enhance performance over the past decade, are illustrated in Fig. 4. Yang et al. proposed an adjacent-point search algorithm extending the dynamic range by iteratively matching distorted spots with calibration references, as illustrated in Fig. 3(a) \cite{yang2024large}. To overcome the limitation of conventional SHWFS, which can only retrieve slope information, He et al. proposed a high-noise frequency-domain extraction algorithm capable of simultaneously extracting both wavefront slope and curvature. This method integrates Fourier demodulation with the frequency-domain signal extraction algorithm (FTFDM) based on the finite difference method (FDM), as shown in Fig. 3(a) \cite{he2022high}. Low-order harmonic components are extracted from the frequency-domain intensity distribution via the Fourier transform, from which the wavefront slope is derived. Subsequently, the curvature information is calculated from the slope matrix using the finite difference method. Simulations under single and combined aberrations demonstrate that reconstruction errors remain within 2$\%$–5$\%$, with the error varying smoothly as noise levels increase, indicating excellent noise tolerance and robustness. On the other hand, current research focuses on deep learning technologies, with neural networks demonstrating breakthrough potential at three levels: 1) Establishing nonlinear mapping models between wavefront gradients and aberrations to improve the robustness of phase recovery \cite{swanson2018wavefront}; 2) Constructing a spot classification network to enable intelligent matching between distorted spots and nominal patterns, enhancing dynamic range adaptability \cite{li2018centroid,medsker2012hybrid}; 3) Developing an end-to-end network architecture to extract wavefront slope and high-order aberration directly features from sensor images ,as shown in Fig. 3(b) \cite{hu2019learning,dubose2020intensity}. For a comprehensive review of neural network algorithms applied to wavefront sensing, see Guo et al. \cite{guo2022adaptive}.

\begin{figure}[ht]
\centering\includegraphics[width=13cm]{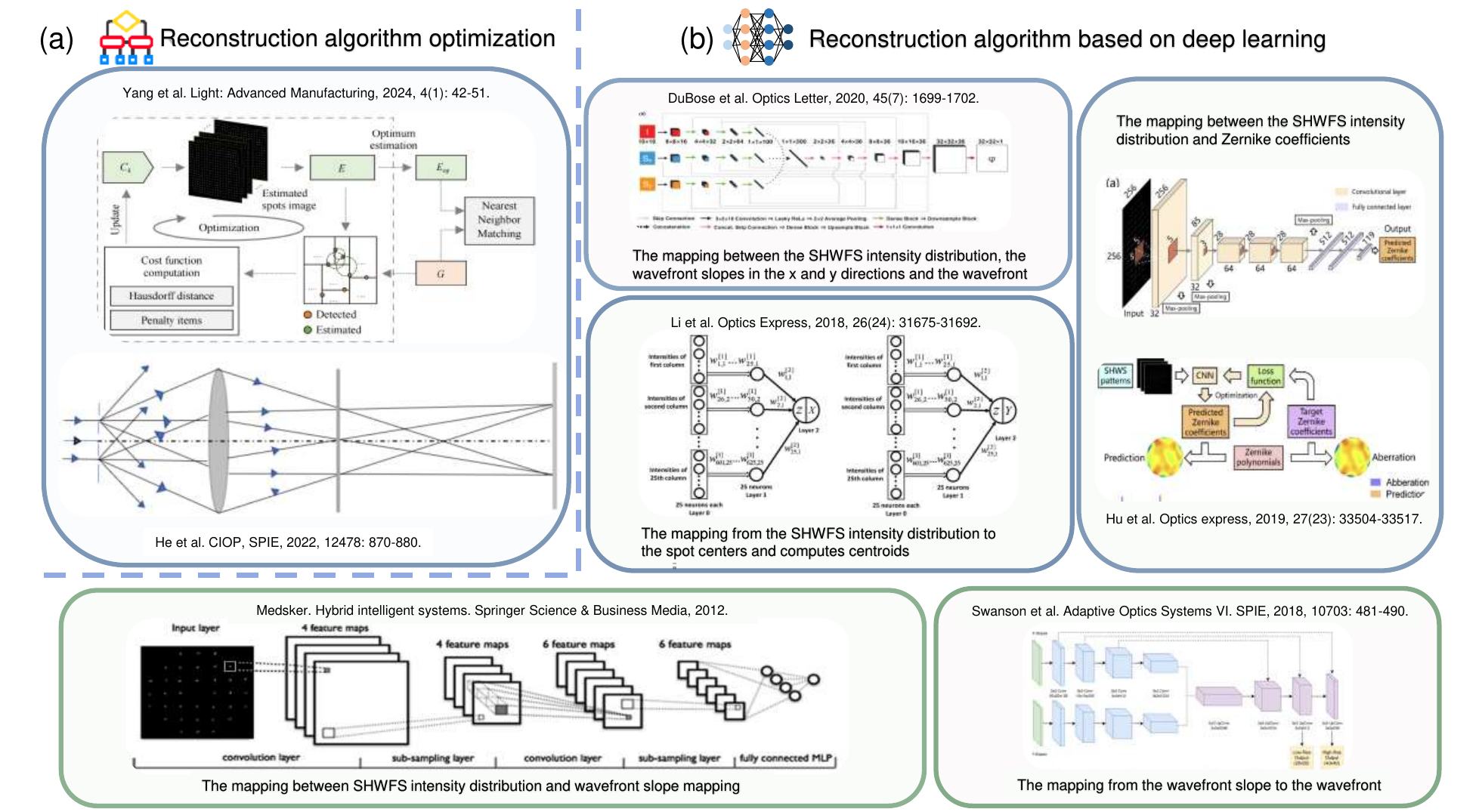}
\caption{A Review of Performance Enhancement Schemes for Algorithm-Based SHWFS. (a) Algorithmic optimization schemes for SHWFS have significantly improved wavefront reconstruction capabilities. One such approach involves a nearest-neighbour search algorithm that rapidly matches distorted spot positions to reference locations, effectively extending the dynamic range of the SHWFS (Ref.\cite{yang2024large}, Fig. 1). Another scheme, known as the Frequency-domain and Finite Difference Method (FTFDM), combines Fourier demodulation with finite difference analysis to overcome the traditional limitation of SHWFS, which typically only provides slope information (Ref.\cite{he2022high}, Fig. 2); (b) Neural network-based wavefront reconstruction has emerged as a powerful alternative, offering several distinct architectures:1) A U-Net-based framework learns the mapping from SHWFS slope data to the true wavefront surface (Ref.\cite{swanson2018wavefront}, Fig. 2). The Shack-Hartmann Neural Network (SHNN) framework first detects spot centers and computes centroids, recasting the spot detection problem as a classification task. It then learns the mapping between distorted and nominal spot positions (Ref.\cite{li2018centroid}, Fig. 3); 2) A convolutional neural network (CNN)-based method uses the full SHWFS intensity image as input, extracting slope information along the optical axis and reconstructing the wavefront (Ref.\cite{hu2019learning}, Fig. 2); 3) The IsNet architecture incorporates three input channels $x$ slope, $y$ slope, and intensity-and directly outputs the corresponding wavefront; 4) The LSHWS model comprises five convolutional layers and three fully connected layers, taking SHWFS intensity images as input and producing Zernike coefficient vectors as output (Ref. \cite{dubose2020intensity}, Fig. 1).}
\end{figure}

The dynamic range limitation of the SHWFS primarily arises from the uncertainty in matching distorted spots to the calibration reference. To address this issue, optical field-shaping technology enhances sensing performance by replacing traditional microlens arrays with diffractive optical elements or SLM to generate spot patterns with spatially differentiated encoding. Building on this concept, in 2015, Sait et al. replaced the microlens array with an optical hologram based on holographic and pattern-matching techniques, assigning distinct pattern structures to each sub-aperture, as illustrated in Fig. 4(a). By employing the correlation peak detection algorithm, the dynamic range of the SHWFS is extended, achieving a measurement accuracy of 0.504 mrad, in contrast to the recovery error of 23.724 mrad observed in conventional Shack-Hartmann sensors \cite{saita2015holographic}. Furthermore, Aftab et al. replaced the microlens array with an LCD screen, as shown in Fig. 4(b). This configuration offers an adaptive dynamic wavefront measurement range capable of accommodating both local and global wavefront changes. The method achieves a high dynamic range measurement of 53.1 mrad \cite{aftab2018adaptive}; On the other hand, the use of diffractive optical elements with greater degrees of modulation flexibility enhances the measurement accuracy of wavefront sensing. In the presence of strong atmospheric turbulence, conventional SHWFS struggle to provide effective measurements. Lechner et al. addressed this challenge by employing a diffractive lens array to measure wavefronts under intense scintillation, adjusting the size of the diffraction region, as shown in Fig. 4(c). The mean residual phase variance was 2.6$\%$ when 45 Zelnik modes were reconstructed \cite{lechner2020adaptable}. To further address the trade-off between the dynamic range and spatial resolution of SHWFS, as well as the limited detection accuracy of thermal blooming wavefronts during high-energy laser atmospheric propagation, Zhang et al. employed a liquid crystal spatial light modulator (LC-SLM) to dynamically generate Fresnel diffractive lens arrays of varying sizes and shapes, replacing conventional fixed-aperture microlens arrays, as shown in Fig. 4(d). Based on the frequency spectral characteristics of the wavefront, a hybrid configuration was designed, featuring a large-aperture lens at the center to enhance the signal-to-noise ratio, and smaller-aperture lenses at the edges to improve sensitivity to high-frequency components. The effectiveness of this approach was experimentally validated under Zernike distortion and simulated thermal blooming conditions \cite{zhang2024shack}. The advancements and key breakthroughs in improved SHWFS technology based on diffractive optical elements over the past decade are summarized in Fig. 4.
\begin{figure}[ht]
\centering\includegraphics[width=12cm]{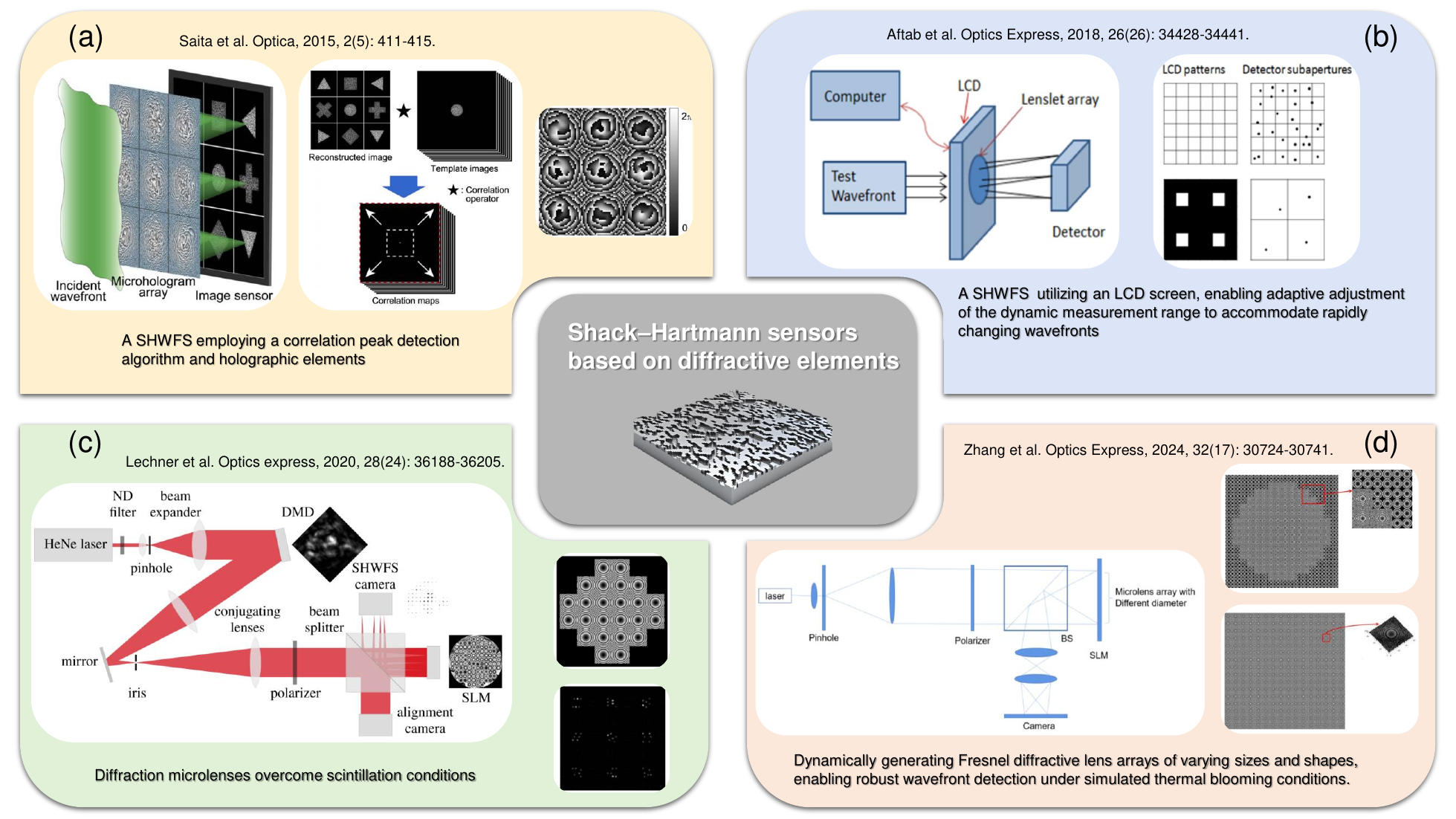}
\caption{Overview of Shack–Hartmann sensors based on diffractive elements. (a) A SHWFS employing a correlation peak detection algorithm and holographic elements (Ref. \cite{saita2015holographic}, Figs. (3)(4)(7)), which retrieves inverse wavefront gradients of various patterns through cross-correlation analysis; (b) A SHWFS  utilizing an LCD screen (Ref. \cite{aftab2018adaptive}, Fig. 2), enabling adaptive adjustment of the dynamic measurement range to accommodate rapidly changing wavefronts; (c) A SHWFS  based on a diffractive lens array (Ref. \cite{lechner2020adaptable}, Fig. 2), capable of measuring wavefronts under strong scintillation conditions by tailoring the size of the diffraction region; (d) A SHWFS  employing a LC-SLM (Ref. \cite{zhang2024shack}, Fig. (2)(3)(9)), dynamically generating Fresnel diffractive lens arrays of varying sizes and shapes, thereby replacing conventional fixed-aperture microlens arrays and enabling robust wavefront detection under simulated thermal blooming conditions.}
\end{figure}

Diffractive optical elements (DOEs) and SLM enable more complex light field manipulation, pushing beyond the performance limits of traditional wavefront sensors and facilitating compact designs. This advancement has inspired significant improvements in conventional solutions such as SHWFS. The phase encoding of incident light wavefronts through microstructural design has driven innovations in light field control modules for wavefront sensing systems. As an extension of DOEs at the subwavelength scale, metasurfaces not only inherit the wavefront modulation properties of DOEs but also overcome the limitations of diffraction efficiency and design flexibility by enabling independent control of multiple parameters, such as phase, amplitude, and polarization, within the subwavelength unit structure. In the following section, we review the application of metasurfaces in SHWFS, offering a valuable reference for researchers in this field.

A distinguishing feature of metasurfaces, compared to conventional diffractive optical elements, lies in the ability of their subwavelength-scale meta-atoms to exhibit polarization-dependent responses. In 2018, Yang et al. integrated a Shack–Hartmann wavefront sensor with a polarization-sensitive metasurface to enable simultaneous detection of both wavefront and polarization information \cite{yang2018generalized}. The optical modulation component of the system comprised six distinct arrays of transmissive meta-lenses (Fig. 5(a). By varying the dimensions and orientation angles of elliptical silicon nanopillars (Fig. 5(b)), the metasurface enabled control over both the transmitted intensity and phase (Fig. 5(c)) of the light field. The scheme encodes the phase and polarization of the light field using a combination of geometric-phase and propagation-phase metasurfaces. Each metalens is engineered to focus light on a specific polarization state, linear polarizations at 0°, 45°, 90° and 135 °, as well as circular polarizations of the left and right. By measuring the focal intensities corresponding to these six polarization channels, the four Stokes parameters could be reconstructed, thereby revealing the complete polarization state of the incident light (Figs. 5(d)~(f)). However, acquiring four Stokes parameters from six separate measurements introduces redundancy, compromising spatial resolution. To address this, the same group proposed in 2020 a redesigned metasurface consisting of four sub-arrays, optimized to enhance both spatial resolution and focusing efficiency \cite{wang2020dielectric} (Fig. 6). This new design achieved a 1.5-fold improvement in spatial resolution. It increased the average focus efficiency from 26$\%$ to 55$\%$.
\begin{figure}[ht]
\centering\includegraphics[width=12cm]{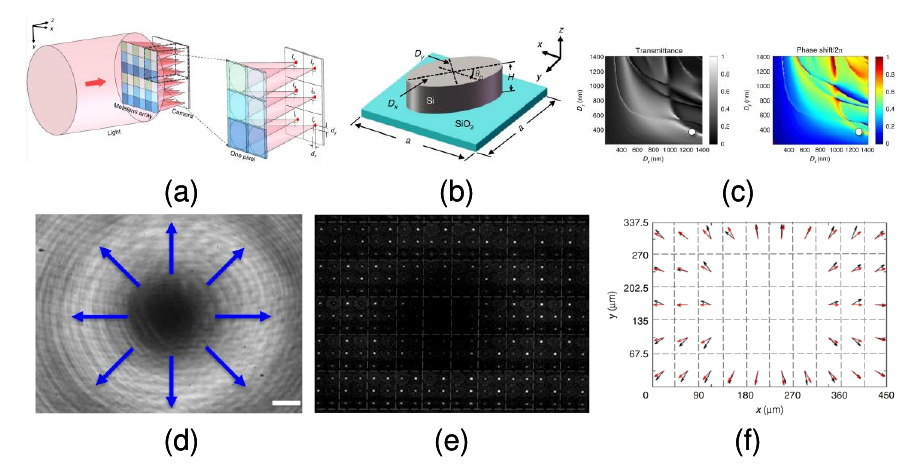}
\caption{Shack–Hartmann Wavefront and Polarization Sensor Based on a Polarization-Dependent Metasurface Array. (a) Schematic of the Shack–Hartmann wavefront sensor utilizing polarization correlation (Ref.\cite{yang2018generalized}, Fig.1); (b) Structural diagram of the meta-atom unit(Ref.\cite{yang2018generalized}, Fig.2(a)); (c) Transmission and phase distributions correspond to different superatomic array sizes(Ref.\cite{yang2018generalized}, Fig.2); (d) Phase and polarization distribution of the incident light field(Ref.\cite{yang2018generalized}, Fig.4(a)); (e) Intensity distribution was measured by the metasurface Shack–Hartmann sensor(Ref.\cite{yang2018generalized}, Fig.4(c)); (f) Reconstructed polarization distribution of the light field(Ref.\cite{yang2018generalized}, Fig.4(e)).}
\end{figure}

\begin{figure}[ht]
\centering\includegraphics[width=12cm]{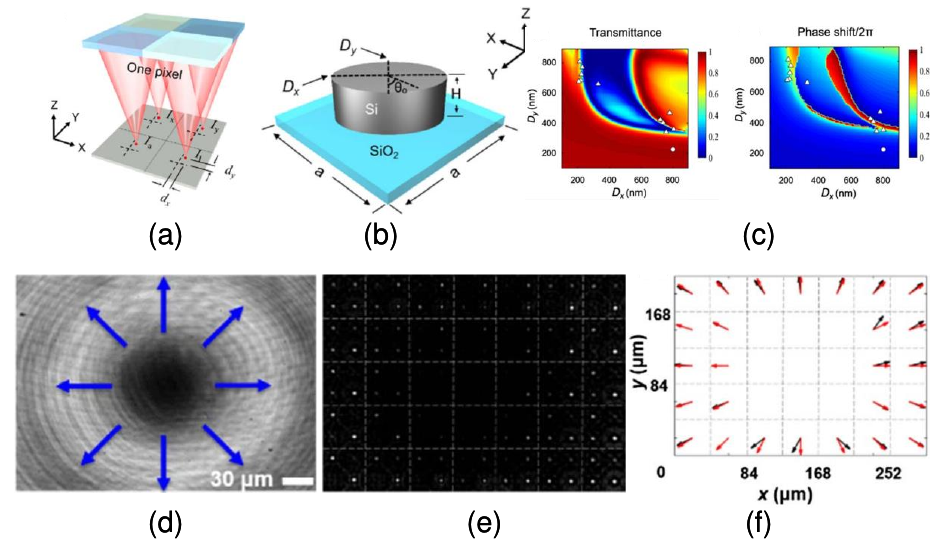}
\caption{Optimized Shack–Hartmann multiparameter detection system based on a polarization-dependent metasurface array. (a) Schematic of the Shack–Hartmann detection system employing polarization correlation (Ref.\cite{wang2020dielectric}, Fig.2(a)); (b) Structural diagram of the meta-atom unit(Ref. \cite{wang2020dielectric}, Fig.2(b)); (c) Transmittance and phase distribution as functions of superatomic array dimensions(Ref. \cite{wang2020dielectric}, Fig.2(c)(d)); (d) Phase and polarization distribution of the incident light field(Ref. \cite{wang2020dielectric}, Fig.5(a)); (e) Intensity distribution measured by the metasurface-based Shack–Hartmann sensor(Ref. \cite{wang2020dielectric}, Fig.5(c));  (f) Reconstructed polarization distribution of the light field(Ref. \cite{wang2020dielectric}, Fig.5(e)).}
\end{figure}

Metasurface-based SHWFS offers a promising pathway toward complex amplitude sensing of optical fields, and recent studies have reported substantial advances in SHWFS architecture. In 2024, Go et al. introduced a meta–shack–Hartmann complex amplitude sensor by replacing conventional microlens arrays with hyperlens arrays, enabling enhanced light field modulation capabilities \cite{go2024meta}, as illustrated in Fig. 7. According to Shannon's sampling theorem, accurate phase sampling requires a sampling frequency at least twice that of the maximum phase gradient. Traditional microlens arrays are constrained by a trade-off between lens density and focal curvature, limiting their ability to detect rapidly varying wavefronts and restricting phase sampling in regions with large gradients. The subwavelength structure of the hyperlens overcomes this limitation by enabling significantly higher sampling densities within the same area (Fig. 7(a)), thereby allowing accurate measurements even in the presence of steep phase gradients.

The operating principle of the metasurface-enabled Shack-Hartmann sensor remains rooted in tracking spot displacements to infer wavefront distortions, as shown in Fig. 7(b). By systematically analyzing the maximum allowable displacement $\Delta_{max}$, positional resolution $\Delta_{res}$, and their ratio $\Delta_{max}$ /$\Delta_{res}$ across varying focal lengths, the authors identified optimal performance at a focal length of 30 \textmu m (Fig. 7(c)). Furthermore, the metalens design was optimized based on criteria such as maximum dynamic range, the number of resolvable angular positions, phase characteristics, and the geometry of the meta-atom structures, thereby further enhancing device performance. Ultimately, a meta-Shack-Hartmann sensor comprising a 100 × 100 metalens array was realized, with each metalens having a diameter of 12.95 \textmu m, a focal length of 30 \textmu m, and a numerical aperture (NA) of 0.21 (Fig. 7(d)). This integrated metasurface array achieves exceptional sampling performance and high-resolution wavefront reconstruction. Using the device as a testbed, Go et al. conducted a series of phase retrieval experiments (Fig. 7(e)), demonstrating that the proposed wavefront sensing scheme offers enhanced spatial resolution and an improved capability to measure large phase gradients. Notably, for phase imaging in the peripheral regions of the image, the error is relatively large, approximately 0.12$\lambda$, which aligns with imaging theory. Specifically, the edge regions exhibit significant aberrations, making imaging more challenging. Additionally, phase imaging based on wavefront sensing imposes certain requirements on the coherence of the light source. In the case of incoherent light, the phase across different regions of the wavefront fluctuates randomly. According to the Van Citter-Zernike theorem, the relationship between the light source information and the coherent region is expressed as $D_{coh }$=4 $\lambda L$ / $\pi D_{led}$, where $D_{coh}$ represents the size of the coherent region within which the wavefront can be considered spatially coherent, $L$ is the distance from the LED to the metal lens array, and $D_led$ denotes the size of the LED.
\begin{figure}[ht]
\centering\includegraphics[width=12cm]{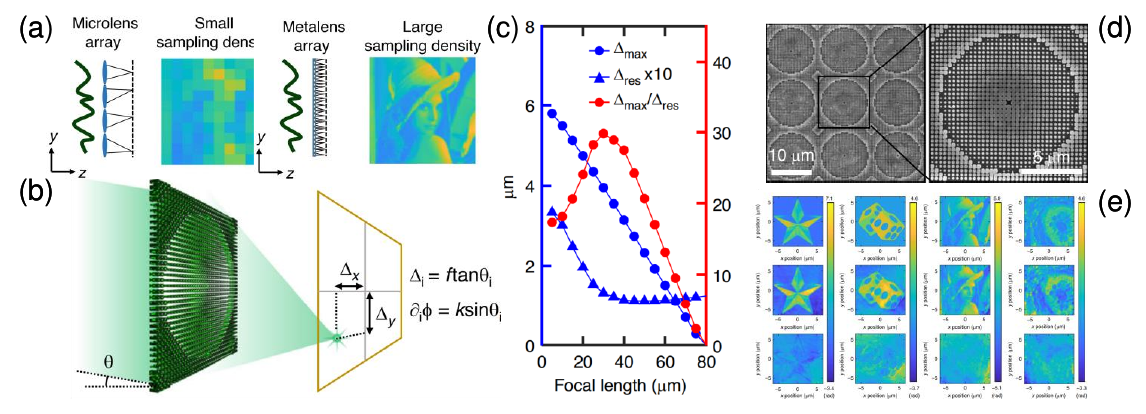}
\caption{Shack–Hartmann wavefront sensor based on metalens. (a) Comparison between a conventional microlens array-based SHWFS and a metalens-based variant. The metalens design offers higher spatial sampling frequency, enabling more accurate detection of steep phase gradients (Ref. \cite{go2024meta}, Fig. 1(a)); (b) Schematic of the metalens based SHWFS scheme, which, akin to the microlens array, encodes phase information via the displacement of focal spots (Ref. \cite{go2024meta}, Fig. 1(a)); (c) Maximum allowable displacement ($\Delta_{max}$), positioning accuracy ($\Delta_{res}$), and their ratio ($\Delta_{max}$/$\Delta_{res}$) across different focal lengths. The optimal performance is observed at a focal length of 30 \textmu m, where $\Delta_{max}$/$\Delta_{res}$ reaches its peak (Ref. \cite{go2024meta}, Fig. 1(d)); (d) Scanning electron microscope image of the fabricated metalens array (Ref. \cite{go2024meta}, Fig. 2(a)); (e) Phase retrieval results. The first, second, and third rows correspond to the phase truth image, the reconstructed phase, and their differences, respectively (Ref. \cite{go2024meta}, Fig. 4(b)).}
\end{figure}

\subsection{Wavefront inversion under multiple intensity constraints: synergistic curvature sensing, phase diversity, and the adaptive potential of metasurfaces} 
Phase information of a light field is partially encoded in the intensity distribution recorded by a detector. In the context of wavefront sensing, this intensity data can be used for phase reconstruction. However, the phase content embedded in a single intensity distribution is inherently limited, making it insufficient for accurate wavefront retrieval. To overcome this, multiple intensity measurements are typically acquired—either through optical field modulation or by translating the image plane—to impose additional constraints on the phase. The wavefront is then reconstructed using numerical iteration or neural network-based algorithms. A comprehensive overview of recent advances in phase imaging using neural networks and multi-intensity distributions can be found in the review by Wang et al. (2024) \cite{wu2019wish}. Curvature wavefront sensing is a representative technique that employs such multi-plane intensity distributions to achieve phase recovery.

The fundamental distinction between the curvature wavefront sensor and the Hartmann sensor lies in the method of converting the local gradient vector field of the wavefront into a local curvature scalar field. This characteristic provides a key advantage in astronomical observations: since atmospheric turbulence has a significantly greater impact on local wavefront slopes than on local curvatures [by Roddier, 1988], the curvature wavefront sensor demonstrates superior robustness in reconstructing distorted wavefronts. A conventional curvature wavefront sensor typically comprises a converging lens and two symmetrically positioned image sensors. The operating principle relies on the intensity variation in the plane perpendicular to the optical axis during light propagation. When an undistorted plane wave is incident, the intensity distributions recorded by the two detectors exhibit uniform symmetry. However, when the wavefront is distorted, asymmetrical intensity deformations appear on both sides. The local wavefront curvature can be determined by analyzing the intensity differences at corresponding positions [by Roddier, 1988]. Its mathematical expression is given as follows:
\begin{eqnarray}
\left.\frac{I_{2}(\mathbf{r})-I_{1}(\mathbf{r})}{I_{2}(\mathbf{r})+I_{1}(\mathbf{r})}=\frac{f(f-l)}{2 l}\left[\frac{\partial}{\partial n} z(f \mathbf{r} / l) \delta_{c}+\nabla^{2} z(f \mathbf{r} / l)\right)\right],
\end{eqnarray}
where $I_1$ and $I_2$ represent the light intensity distributions on the two image sensors, respectively. $\delta_c$ denotes the linear pulse distribution at the pupil edge, weighted by the radial tilt of the wavefront $\delta z$/$\delta n$. $\nabla^{2}z$ corresponds to the wavefront curvature, f is the focal length of the focusing optical system, and $l$ represents the distance between the detector and the focal plane. $\mathbf{r}=\sqrt{x^{2}+y^{2}}$ further denotes different positions within the wavefront distribution.

The core advantages of curvature wavefront sensors include: (1) the ability to detect high-order aberrations within a low dynamic range; and (2) a direct mapping relationship between the sensing signal and the curvature correction of a deformable mirror, making them highly suitable for fast adaptive optical systems. This technique was initially applied to biological eye wavefront detection by Cristiano and Fernando's team and has since been extended to astronomical observations \cite{torti2008wavefront,diaz2006curvature}. Notably, its fast response characteristics have led to its adoption in wavefront sensing systems for large-scale telescopes, such as the Large Synoptic Survey Telescope (LSST) \cite{xin2015curvature}. According to aberration theory, high-order aberrations result in an asymmetric wavefront distribution, where the local radial curvature is influenced by both the $x$ and $y$ directions. The conventional approach of directly estimating radial curvature using a curvature wavefront sensor inherently introduces errors. To overcome this limitation, Manuel et al. proposed an improved solution: replacing the converging lens with a 4f optical system and inserting an amplitude-encoding mask at the intermediate focal plane, as illustrated in Fig. 8(c) \cite{cagigal2015x}. By measuring the intensity distribution differences of the beam modulated by different encoding patterns on the CCD, the wavefront curvature components in the $x$ and $y$ directions can be reconstructed independently, significantly enhancing measurement accuracy. Nonlinear wavefront sensing technology has emerged to overcome the performance limitations of traditional curvature sensing. This approach integrates the discrete sampling concept of the Shack-Hartmann sensor with the phase retrieval mechanism of curvature sensing. The core optical configuration is as follows: First, the pupil plane is spatially sampled, with each subregion beam passing through an independent optical chain consisting of defocusing elements and converging lenses. These beams then undergo overlapping interference on the detector surface. Sub-nanometer-scale wavefront curvature measurement is achieved by analyzing the coherent phase information encoded in the interference pattern and applying a nonlinear reconstruction algorithm \cite{guyon2009high}. Experiments demonstrate that this technology enhances measurement sensitivity by order of magnitude and has been successfully implemented to upgrade the adaptive optics system of the Subaru Telescope \cite{ahn2023development}. The technological advancements and cutting-edge applications of curvature wavefront sensors are illustrated in Fig. 8. 
\begin{figure}[ht]
\centering\includegraphics[width=12cm]{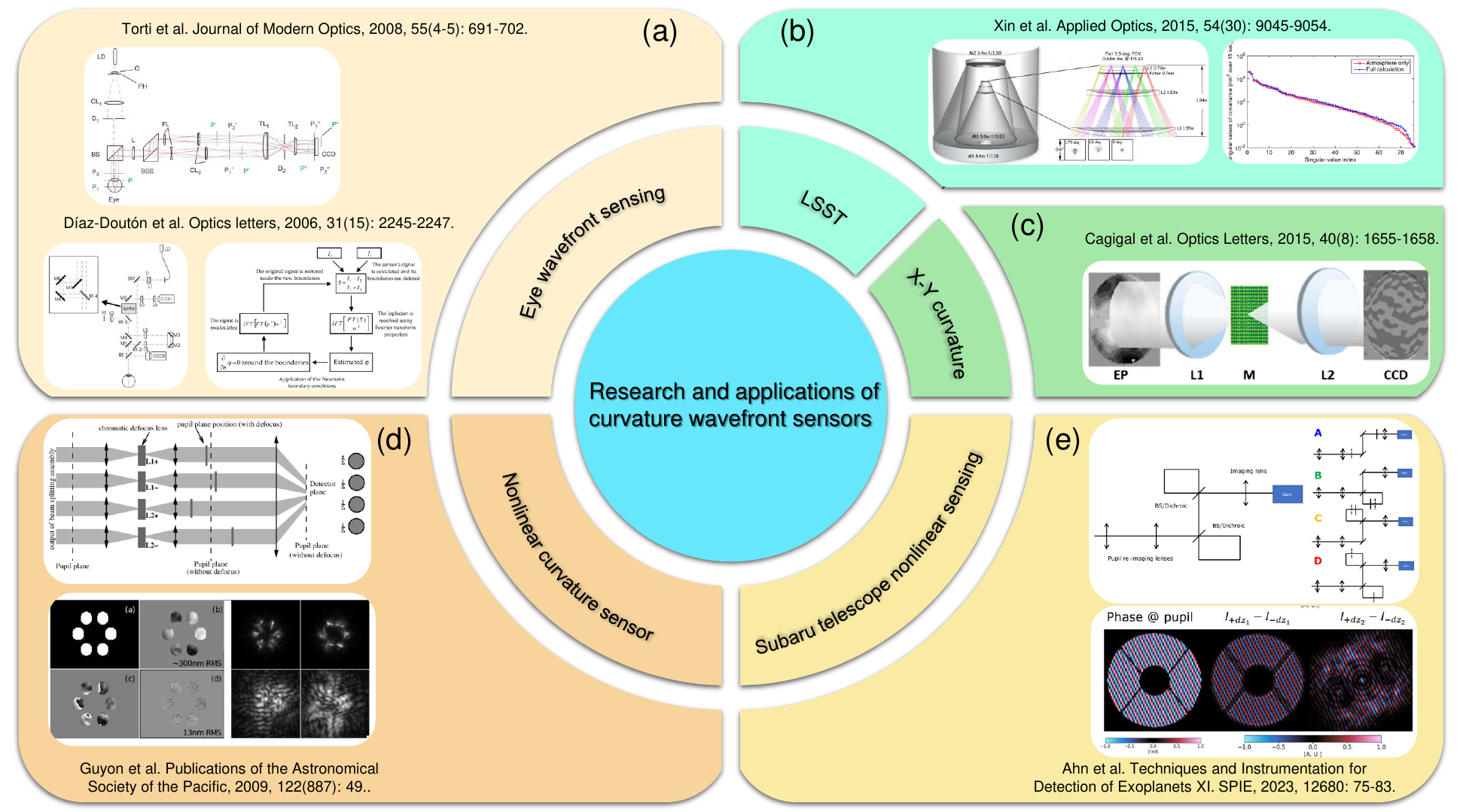}
\caption{Overview of research and applications of curvature wavefront sensors. (a) Biological eye wavefront sensing based on curvature wavefront sensors (Ref. \cite{torti2008wavefront}, Fig. 1)(Ref. \cite{diaz2006curvature}, Fig. (1)(2)); (b) Large survey telescopes employing curvature wavefront sensors (Ref. \cite{xin2015curvature}, Figs. (1)(10)); (c) Curvature wavefront sensing scheme based on amplitude coding in the frequency domain, where the intermediate image plane of a 4$f$ optical system is encoded, and curvature demodulation in the $x$ and $y$ directions is performed using intensity distributions modulated by distinct coding patterns (Ref. \cite{cagigal2015x}, Fig. 1); (d) Nonlinear curvature wavefront sensor, in which spatial sampling is performed at the pupil plane, beams from each sub-region travel through independent optical paths, and overlapping interference occurs at the detector plane, allowing phase information to be demodulated from the interference fringes (Ref. \cite{guyon2009high}, Fig. 1); (e) Nonlinear wavefront sensors applied in the Subaru telescope system (Ref. \cite{ahn2023development}, Figs. (2)(3)).}
\end{figure}

The curvature wavefront sensor acquires multiple intensity distributions by physically shifting the image plane, enabling phase retrieval through light field propagation. To date, this approach has not been integrated with metasurface technology. In contrast, an alternative phase retrieval method, based on optical field coding to generate multiple intensity distributions, has been successfully combined with metasurfaces. This scheme modulates the incident light field using a series of known and mutually uncorrelated patterns, producing corresponding intensity distributions at the detector. The wavefront phase is then reconstructed from this set of intensity measurements. In 2019, Wu et al. \cite{wu2019wish} proposed the use of a SLM to load eight uncorrelated random phase patterns, enabling wavefront phase reconstruction via an optimized iterative algorithm. Building on this approach, they conducted phase retrieval experiments involving handprints and scattering media, all of which demonstrated excellent performance, as illustrated in Fig. 9.
\begin{figure}[ht]
\centering\includegraphics[width=12cm]{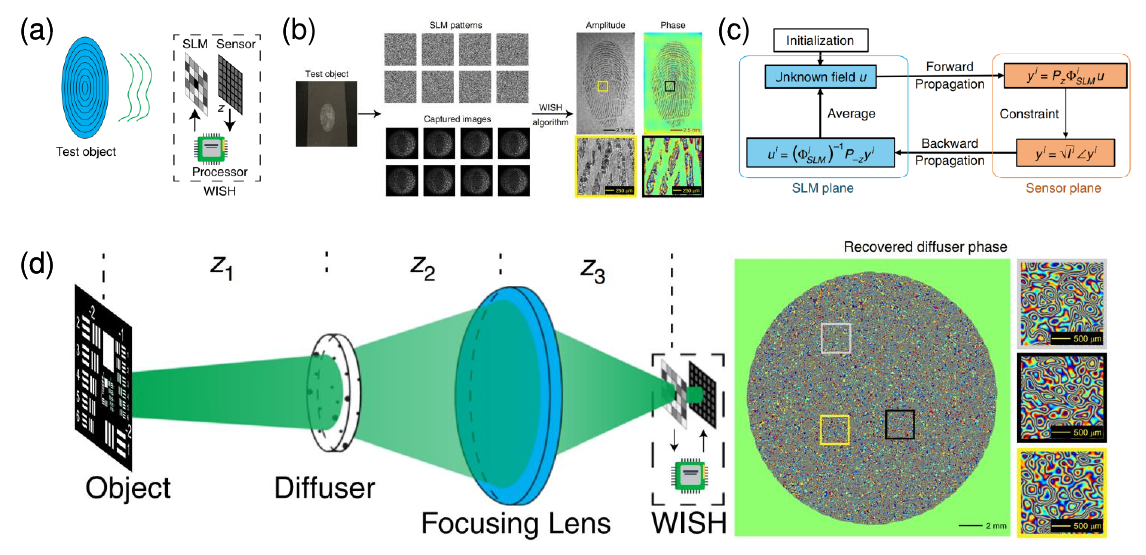}
\caption{Principle of phase diversity schemes based on multi-intensity distributions. (a) Hardware configuration of the phase diversity system, comprising a SLM, detector, and processor (Ref. \cite{wu2019wish} Fig. 1(a)); (b) Core principle: By applying multiple uncorrelated phase patterns to the SLM, a series of corresponding intensity distributions are captured at the detector. These measurements serve as constraints for regularized optimization of the unknown incident phase (Ref. \cite{wu2019wish} Fig. 2(c)); (c) Phase retrieval process based on the Gerchberg–Saxton (GS) algorithm (Ref. \cite{wu2019wish} Fig. 2(b)); (d) Reconstruction of the phase distribution of a scattering medium using phase diversity, including the optical setup and corresponding results (Ref. \cite{wu2019wish} Fig. 4(b)(c)).}
\end{figure}

However, this approach requires sequential loading of multiple uncorrelated modulations onto the SLM, resulting in low sensing efficiency. In contrast, metasurfaces offer a high degree of freedom in light field modulation, enabling simultaneous encoding of diverse optical information. This intrinsic capability positions metasurfaces as a promising candidate for integration with phase diversity schemes. Researchers leverage the polarization sensitivity of the metasurface to decompose the in-ray polarized light into two orthogonal circular polarization components, generating an intensity distribution across different regions of space. This design enables the simultaneous acquisition of multiple-intensity data through a single exposure, thus avoiding the alignment errors introduced by repeated focusing and enhancing the robustness of phase reconstruction. In 2024, Jimenez et al. \cite{jimenez2024single} used a birefringent metasurface to convert the axial PSF into eight independent PSF distributions within the same focal plane as illustrated in Fig. 10. These PSFs are separated by dual channels of spatial position and polarization state (left/right circular polarization). The mapping relationship between the PSF set and the phase to be measured is directly established using a U-Net neural network, and the effectiveness of this end-to-end phase recovery method is validated through experiments. In 2022, Zhou et al. \cite{zhou2022single} proposed a single-lens phase retrieval scheme based on an anisotropic metasurface. The core principle of this method lies in decomposing the energy of incident polarized light into orthogonal circular polarization components, then employing the metasurface to generate three defocused intensity images at distinct coaxial positions. Phase reconstruction is subsequently achieved via an improved three-image transport-of-intensity equation (TTIE) algorithm, as illustrated in Fig. 11. Experimental demonstrations using a ring-shaped phase distribution and a badge-shaped pattern (Fig. 11(c)) validate the feasibility and precision of this approach. Building on this foundation, Zhou et al. extended their work to quantitative phase microscopy imaging. Using the spatial multiplexing capability of the metasurface in combination with deep learning algorithms and the intensity transmission equation, they achieved high-speed, high-precision phase reconstruction, as illustrated in Fig. 12 \cite{zhou2025advanced}. Additionally, wavefront and phase retrieval schemes based on multiplane intensity distributions offer a novel approach for characterizing the intrinsic phase profiles of metasurfaces. The phase modulation properties of the metasurface can be retroactively inferred by analyzing the far-field intensity distribution of a collimated beam after modulation by the metasurface. This method significantly improves the accuracy of metasurface phase detection, as shown in Fig. 13 \cite{liu2024metalenses,cheng2025quantitative}. 
\begin{figure}[ht]
\centering\includegraphics[width=12cm]{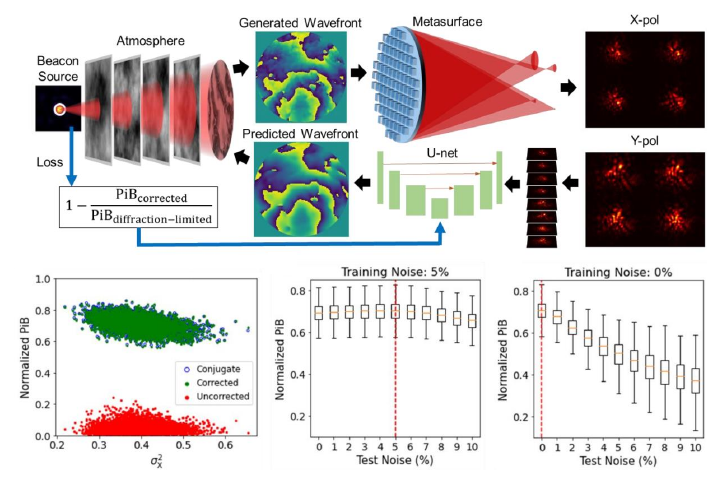}
\caption{Atmospheric turbulence phase reconstruction enabled by birefringent metasurfaces. A birefringent metasurface is employed to divide the point spread function (PSF) within a single focal plane into eight independent distributions, separated by spatial position and polarization channels (left- and right-handed circular polarization). A U-Net neural network is trained to learn the mapping between the PSF ensemble and the corresponding turbulence phase. The results show that the network can accurately reconstruct turbulence phase profiles, including the effects of scintillation, from the measured PSF ensemble, demonstrating the efficacy of this approach (Ref.\cite{jimenez2024single} Fig. 3).}
\end{figure}

\begin{figure}[ht]
\centering\includegraphics[width=12cm]{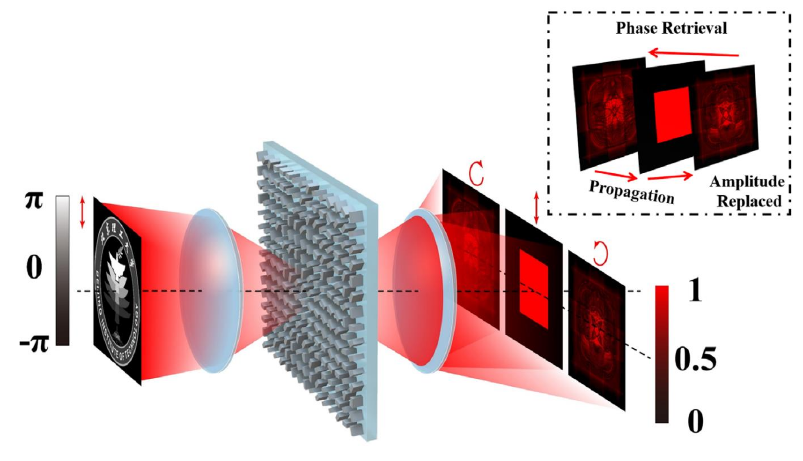}
\caption{Triple-intensity distribution imaging for phase retrieval based on anisotropic metasurface intensity transfer equations. Linearly polarized light waves are modulated by an anisotropic metasurface in the frequency domain. Following inverse Fourier transformation, three parallel intensity distributions are formed and captured in a single exposure: one in focus and two defocused, with fixed and conjugate defocus distances. By combining these three intensity measurements with the intensity transport equation algorithm, quantitative phase images can be accurately reconstructed (Ref. \cite{zhou2022single} Fig. 1).}
\end{figure}

\begin{figure}[ht]
\centering\includegraphics[width=12cm]{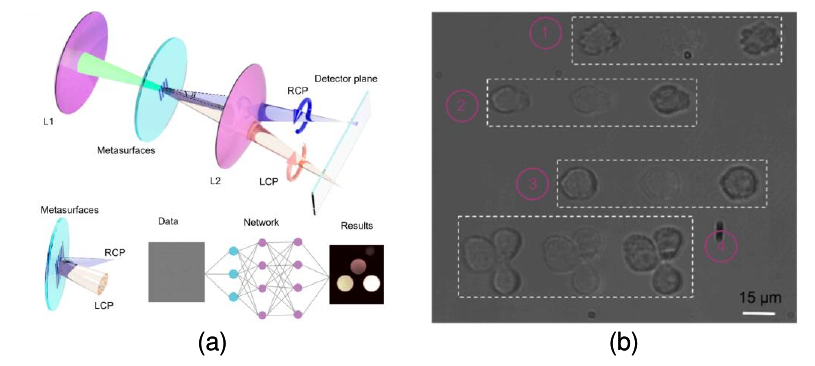}
\caption{Schematic of a quantitative phase microscopy system based on a metasurface. (a) In a 4f optical system, right-circularly polarized light is focused in front of the detector plane, while left-circularly polarized light is focused behind it, generating controlled defocus patterns. The metasurface manipulates left- and right-circularly polarized light, steering them along distinct optical paths. The intensity images captured by the CCD are processed and quantitatively reconstructed using deep learning techniques (Ref. \cite{zhou2025advanced} Fig. 1). (b) Based on this principle, quantitative phase imaging experiments of biological cells were successfully demonstrated. (Ref. \cite{zhou2025advanced} Fig. 4(v)).}
\end{figure}

\begin{figure}[ht]
\centering\includegraphics[width=12cm]{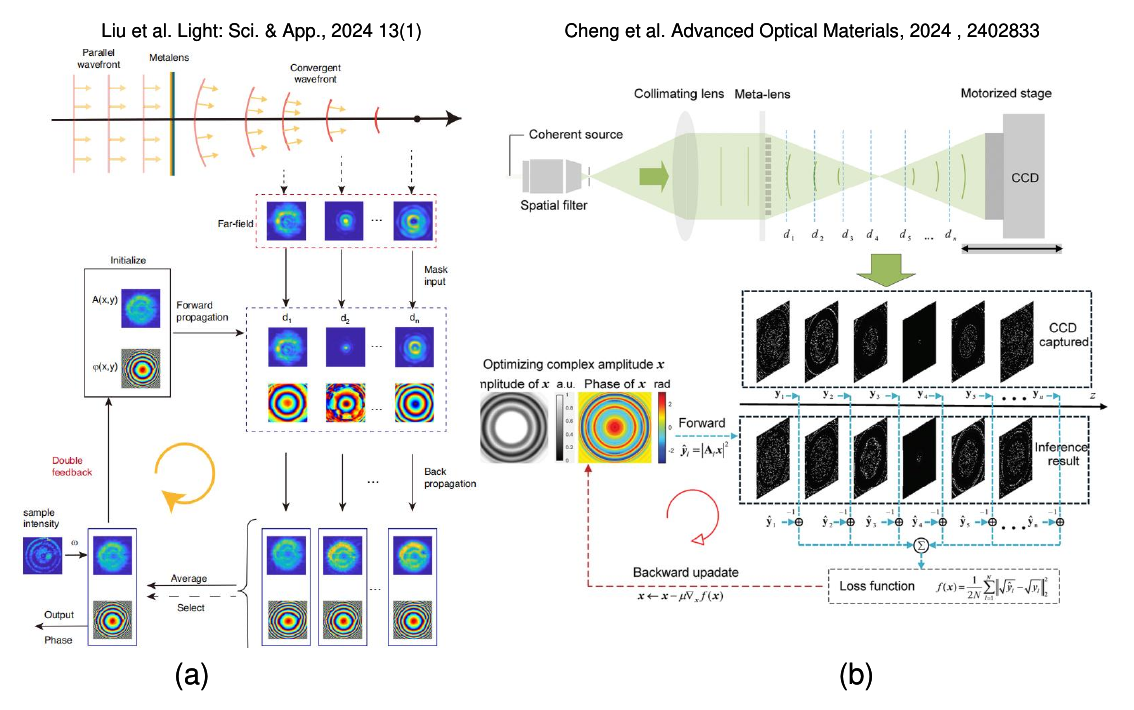}
\caption{Phase retrieval for precise hypersurface characterization. Utilizing a phase retrieval approach, the intrinsic phase profile of the metasurface is reconstructed, enabling precise phase detection. (a) The proposed measurement strategy employs a collimated beam incident on the metasurface, capturing intensity distributions at multiple axial distances. Phase retrieval is performed by leveraging these intensity distributions. The algorithm initiates with a random field distribution and propagates it forward from the sample surface using angular spectrum propagation. The phase distribution is retained, while the computed intensity distribution is replaced with experimentally measured data. The updated far-field distribution is then back-propagated to the sample plane, generating $N$ corresponding field distributions. The average of these $N$ complex amplitude distributions is computed, and the predicted metasurface field is iteratively refined based on the measured intensity $I$ at each position.  Using the averaged sample-plane field distribution, the far-field intensity at $N$ positions is recalculated, and the complex amplitude is updated accordingly. Through iterative propagation, the phase distribution of the metasurface is ultimately reconstructed (Ref. \cite{liu2024metalenses} Fig. 2). (b) An alternative phase retrieval approach utilizes intensity distributions at multiple axial positions.  Here, the deviation between the iteratively retrieved intensity distribution and the target intensity distribution serves as the loss function, enabling phase recovery via gradient descent optimization (Ref. \cite{cheng2025quantitative} Fig. 1).}
\end{figure}

\subsection{Subwavelength Reconstruction of Wavefront Shear Interference: From Classical Grating to Metasurface-Enabled Control}

Methods based on multi-intensity distribution provide a simple and direct approach for wavefront sensing and phase retrieval, offering a promising direction for innovation in the field. However, extracting phase information embedded within intensity measurements often lacks analytical solutions, necessitating iterative optimization through non-convex algorithms, typically gradient-based. Even with the adoption of advanced techniques such as automatic differentiation or analytic gradient approximations, the overall computational speed remains limited. Another one, that neural network–based approaches offer significantly faster inference but often require extensive data training to achieve an accuracy comparable to iterative methods. As such, current multi-intensity-based approaches face inherent trade-offs between computational speed and retrieval accuracy.

Light interference arises fundamentally from variations in the optical phase, and thus, changes in interference fringes directly encode phase shifts in the light field. Phase and wavefront information can therefore be efficiently retrieved by analyzing the interference pattern. Owing to the high sensitivity of interference fringes to minute phase variations, this method offers exceptional precision in wavefront sensing and phase recovery. Furthermore, it typically does not require iterative computation, enabling rapid phase reconstruction. Among classical approaches, shear interferometry stands out as a widely adopted and robust technique in the domain of wavefront sensing.

The core principle of interference-based wavefront sensing lies in decomposing the incident beam into two spatially offset, coherent light fields via a beam-splitting element, thereby generating interference fringes on the detector plane. By analyzing the phase and deformation characteristics of these fringes, the spatial distribution of the incident wavefront can be reconstructed. In practical engineering implementations, the four-wave shear interferometer utilizes a two-dimensional complex amplitude grating as the core splitting element. This grating can selectively suppress even and triple diffraction orders (while preserving the $\pm 1$ diffraction orders) through precision design, significantly improving the signal-to-noise ratio of the interference signal \cite{primot2000extended}. This unique characteristic allows it to maintain high-precision wavefront reconstruction capabilities even in environments with strong noise.

When the light to be measured is diffracted through the grating, four beams of first-order diffracted light are primarily formed. These four beams of first-order diffracted light coherently superimpose on the detector, creating interference fringes. The expression for the interference pattern can be written as follows:
\begin{eqnarray}
\left.I(x, y)=R I_{0}(x, y) \left\lvert\, \begin{array}{l}
1+\frac{1}{2} \cos \left(s k \frac{\partial W(x, y)}{\partial x}+\frac{2 \pi}{d} x\right)+\frac{1}{2} \cos \left(s k \frac{\partial W(x, y)}{\partial y}+\frac{2 \pi}{d} y\right) \\
+\frac{1}{4} \cos \left(\sqrt{2} s k \frac{\partial W(x, y)}{\partial(x+y)}+\frac{2 \pi}{d}(x+y)\right)+\frac{1}{4} \cos \left(\sqrt{2} s k \frac{\partial W(x, y)}{\partial(x-y)}+\frac{2 \pi}{d}(x-y)\right)
\end{array}\right.\right],
\end{eqnarray}
where $W$($x$,$y$) represents the wavefront distribution of the light to be measured, $R$ is the grating diffraction efficiency, $I_0$($x$,$y$)=$A^2$($x$,$y$) is the intensity distribution of the light, $S$ is the shear amount of the four beams of first-order diffracted light along the $x$ or $y$ direction, and $\delta W$($x$,$y$)/$\delta x$ and $\delta W$($x$,$y$)/ $\delta y$ are the wavefront slopes in the x and y directions, respectively. Additionally, $\delta W$($x$,$y$)/ $\delta (x+y)$ and $\delta W$($x$,$y$)/ $\delta (x-y)$ represent the wavefront slopes along the angle between the $x$ and $y$ axes. After performing a Fourier transform of the detector intensity distribution, five frequency spectrum components will emerge, which can be expressed as follows:
\begin{eqnarray}
H_{0}=F T\left(I_{0}(x, y)\right),
\end{eqnarray}
\begin{eqnarray}
H_{ \pm 1}=F T\left(\frac{1}{4} I_{0}(x, y) \exp \left( \pm i\left(s k \frac{\partial W(x, y)}{\partial x}+\frac{2 \pi}{d} x\right)\right)\right),
\end{eqnarray}
\begin{eqnarray}
H_{ \pm 2}=F T\left(\frac{1}{4} I_{0}(x, y) \exp \left( \pm i\left(\operatorname{sk} \frac{\partial W(x, y)}{\partial y}+\frac{2 \pi}{d} y\right)\right)\right),
\end{eqnarray}
where $FT(\cdot)$ denotes the Fourier transform. By selecting an appropriately shaped frequency-domain window, extracting the zero-frequency component $H_0$, and applying the inverse Fourier transform, the intensity distribution $I_0$ can be obtained. Simultaneously, by adjusting the frequency-domain window to isolate the spectral components $H_{\pm 1}$ and $H_{\pm 2}$, shifting them to the center of the frequency domain, and performing an inverse Fourier transform, the wavefront slopes in the $x$ and $y$ directions can be accurately reconstructed.

The measurement accuracy of the four-wave transverse shear interferometer is largely determined by the optimization of the complex amplitude grating’s diffraction efficiency, which must concentrate the diffraction energy into the $\pm 1$ orders while suppressing higher-order stray light interference. The key technological advancements are as follows: In 2000, the Primot team introduced a compound grating structure by superimposing a Hartmann mask onto a checkerboard-phase grating with a phase period of 2$d$ (alternating 0/$\phi$ distribution), enabling maximal energy extraction for the four interfering beams \cite{primot2000extended}. In 2005, Chanteloup et al. demonstrated through frequency-domain theoretical analysis that the transmittance function of an ideal complex amplitude grating should follow a cosine absolute value distribution, providing a theoretical foundation for subsequent randomly coded grating designs \cite{chanteloup2005multiple}. In 2008, Jasoh et al. combined a shear‑interferometric wavefront sensor with a SLM, creating a shear interferometer whose shear ratio can be tuned in real time by modifying the SLM phase pattern \cite{karp2008integrated}. In 2015, Ling et al. proposed a random coding algorithm based on luminous flux constraints. They developed a hybrid structure combining a binarized amplitude grating with a cosine-distributed or checkerboard-phase grating. This design reduced the high-order diffraction energy to just 1.55$\%$ of the $\pm 1$ order energy while enabling a continuously tunable dynamic range \cite{ling2015general,ling2015quadriwave}, as illustrated in Fig. 14. As four‑wave shear interferometry has matured, it has moved into industrial use, offering valuable tools for phase‑contrast microscopy\cite{zhong2024phase}, aspheric‑surface inspection\cite{zhang2019novel}, aberration testing\cite{agour2022characterization,jiang2016measurement,ling2015general} and other field \cite{munj2023unidirectional,guo2020quantitative,zhang2019ultrafast}. 
\begin{figure}[ht]
\centering\includegraphics[width=12cm]{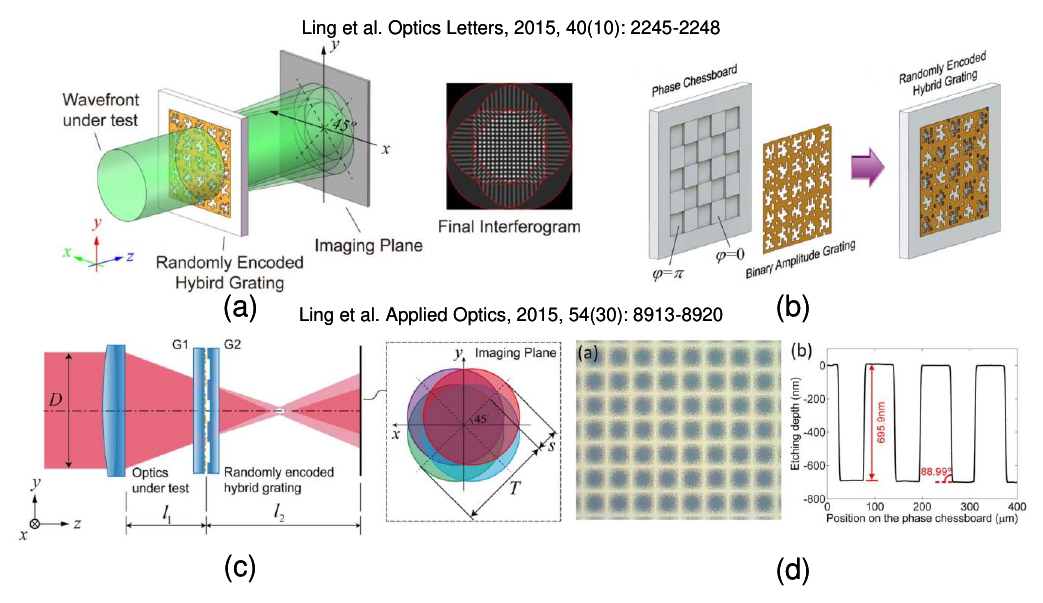}
\caption{Overview of random-coded hybrid grating shear interferometric wavefront sensing.(a) Schematic of the randomly coded raster wavefront sensing setup (Ref. \cite{ling2015quadriwave}, Fig. 1);(b) Structural diagram of the randomly coded raster (Ref. \cite{ling2015quadriwave}, Fig. 2); (c) Layout of the aberration testing system based on randomly coded gratings, where G1 is a randomly coded binary amplitude grating and G2 is a phase checkerboard grating (Ref. \cite{ling2015general}, Fig. 6); (d) Micrographs of the random complex amplitude grating and surface profiles of the phase checkerboard grating (Ref. \cite{ling2015general}, Fig. 8).}
\end{figure}

Shear interferometric wavefront sensing, characterized by independent reference wave calibration and sub-nanometer phase sensitivity, has demonstrated unique advantages in short-wavelength detection, including terahertz and $X$-ray applications \cite{agour2022terahertz,liu2018high,nagler2017focal,makita2020double}. Key technological advancements have centered on the optimal design of spectral gratings to suppress stray light or incorporate additional information, thereby enhancing the signal-to-noise ratio of the measurement. For instance, Goldberg designed a binary amplitude reflection grating that integrates a two-dimensional shear interference grating with a Hartmann mask, combining the advantages of shear interferometry and Hartmann wavefront sensing to enhance both the dynamic range and sensitivity of wavefront measurement \cite{goldberg2020reflective,goldberg2021binary}. Makita further advanced the field by proposing a dual-grating shear interferometry approach for the characterization of X-ray free-electron laser beams \cite{makita2020double}. In this approach, two gratings are positioned along the measurement optical path, with one grating capable of rotation to enable spatial resolution adjustment and tunable wavefront sensitivity. The high accuracy of shear-based wavefront sensing techniques has been widely validated through their extensive applications in laser beam quality assessment and optical surface testing \cite{song2024extended,peng2020calibration,lam2018complete,song2022surface}. The research advancements and latest applications of four-wave shear interferometric sensors are illustrated in Fig. 15. 
\begin{figure}[ht]
\centering\includegraphics[width=12cm]{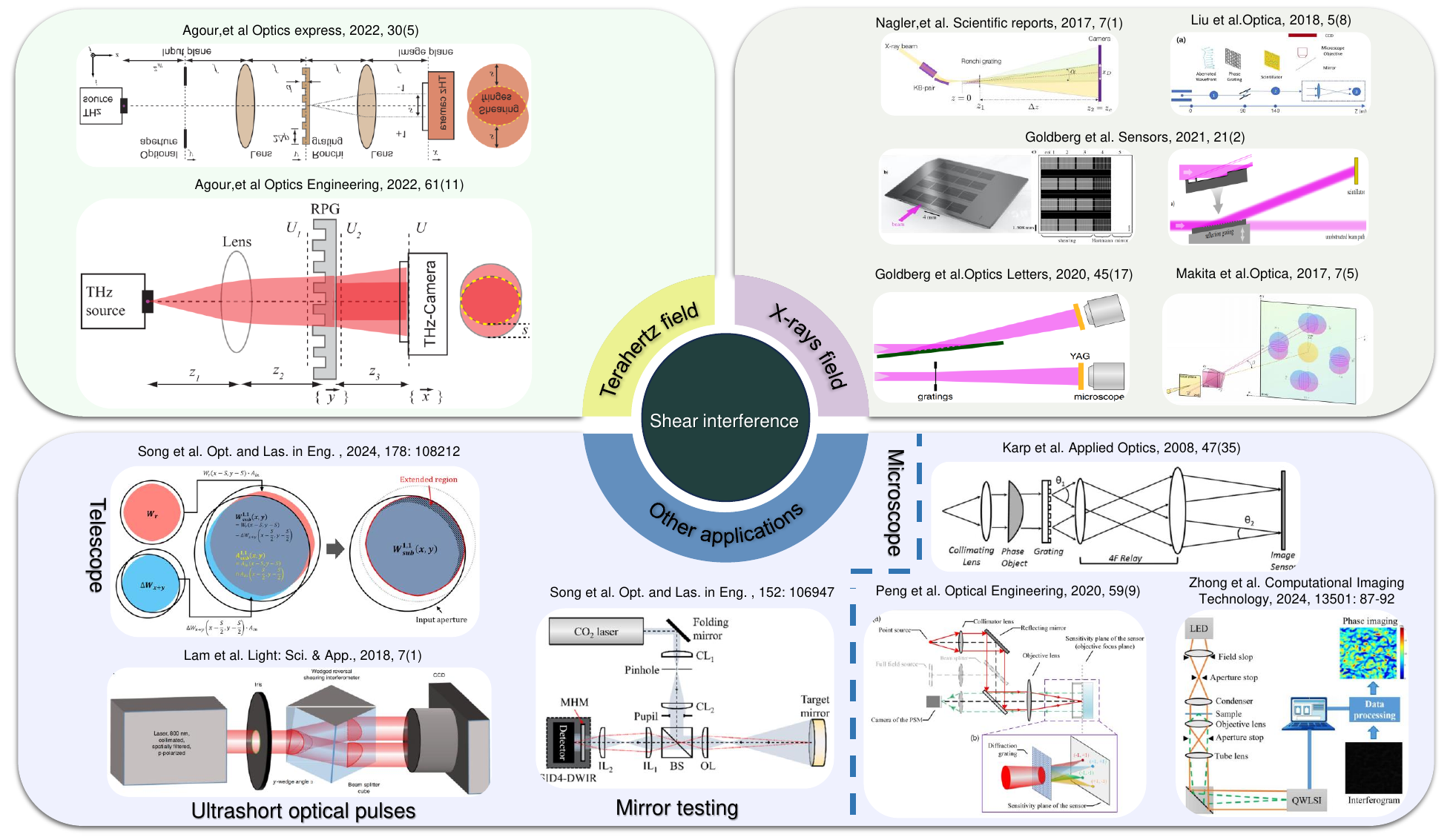}
\caption{Overview of research and applications of shear interferometric wavefront sensing, spanning terahertz, $X$-ray, and other domains.}
\end{figure}

The core of the shear interferometry scheme lies in the ability of a two-dimensional grating to efficiently split the light field, ultimately forming interference fringes on the detector. Traditional approaches optimize the theoretical diffraction efficiency of the two-dimensional grating through analytical design. However, the actual performance of conventional gratings remains limited. By contrast, metasurfaces can achieve high efficiency through precise design of the structure and arrangement of meta-atoms. In addition, their multi-degree-of-freedom tunability offers enhanced control over light propagation. The integration of metasurfaces thus holds great promise for advancing shear interferometry, enabling more compact, efficient, and versatile implementations.

A representative approach to integrating metasurfaces leverages their high modulation capabilities and intrinsic multiplexing properties to enhance the interference of multiple coherent fields, thereby improving the accuracy of wavefront sensing. Wu et al. \cite{wu2023single} proposed a single-point quantitative complex amplitude imaging method based on double all-medium geometric phase hypersurfaces in 2023, as shown in Fig. 16. The core principle of this method is to utilize a pair of PB metasurfaces to encode the amplitude and phase information of an object into a differential interference pattern. Fig. 16(a) illustrates the schematic of the metasurface-based quantitative amplitude and phase imaging (QAPI) system. An $x$ polarized input beam, $E_{in}$$(x,y)$, passes through a polarizer and sequentially illuminates two metasurfaces, $MS_1$ and $MS_2$, positioned at axial distances $z_1$ and $z_2$ from the object, respectively. Each geometric-phase metasurface manipulates the polarization state of the incident beam, converting the left- and right-handed circularly polarized (LCP and RCP) components while imparting geometric phases of $+2\phi_i$ and $-2\phi_i$ to the LCP and RCP beams, respectively. Here $\varphi_{i}=\pi\left(x-\xi_{i}\right) / \Lambda_{i}$,$i$=1, 2 are the designed orientations of the local optical axes of the two metasurfaces, $\Lambda_i$ is the transverse shift of $MS_i$ along $x$ axis, and  is the period of $MS_i$. The resulting phase shifts imparted to the LCP and RCP beams are depicted in Fig. 16(a). Two metasurfaces were used to achieve polarization decomposition of the incident light and beam deflection, generating two transverse displacement light fields of the input object. The two beams of light interfered with a light field with a differential phase delay. A polarization camera was employed to capture four delayed images synchronously, ultimately reconstructing the amplitude and phase information of the object [by Wu et al., Optica, 2023]. Wu et al. performed phase retrieval experiments on biological cells using this scheme, with the reconstructed results presented in Fig. 16(b). This scheme employs the four-step phase-shifting method to calculate the phase gradient in a single direction, expressed as follows:
\begin{eqnarray}
G_{x} \approx \frac{1}{2 \Delta_{0}}\left[\phi\left(x_{3}+\Delta_{0}, y_{3}\right)-\phi\left(x_{3}-\Delta_{0}, y_{3}\right)\right]=\frac{1}{2 \Delta_{0}} \operatorname{atan}\left(\frac{I_{2}-I_{4}}{I_{1}-I_{3}}\right) .
\end{eqnarray}
where $\Delta_0$ denotes the lateral displacement of different intensity pattern, $I_1$, $I_2$, $I_3$ and $I_4$ represent the four shear images acquired at different phase steps.

\begin{figure}[ht]
\centering\includegraphics[width=12cm]{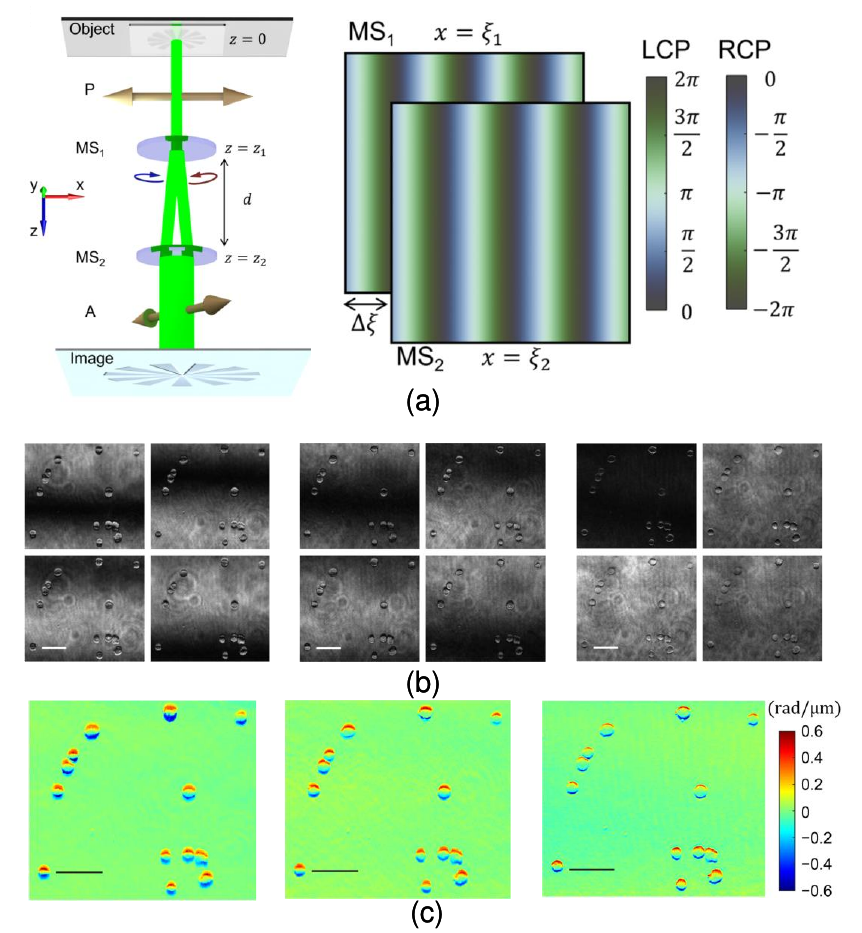}
\caption{A single-point quantitative complex amplitude imaging method based on double-dielectric geometric-phase metasurfaces. (a) Schematic of complex amplitude imaging using two metasurfaces separated by distances $z_1$ and $z_2$. P, polarizer; MS, metasurface; A, analyzer (Ref. \cite{wu2023single} Fig. 1); (b) Simultaneous acquisition of four phase-delay images generated by different pairs of metasurfaces(Ref. \cite{wu2023single} Figs. 3(b)(c)(d)). (c) Quantitative amplitude and phase images of cells reconstructed from the measured phase-delay data (Ref. \cite{wu2023single} Figs. 3(e)(f)(g)).}
\end{figure}

The reconstruction resolution of the quantitative phase image in this approach depends on both the numerical aperture of the objective lens and the lateral displacement between the two interfering light fields, which is determined by the period of the metasurface and its distance from the imaging plane. A smaller lateral displacement improves phase reconstruction accuracy, as it reduces image blurring along the shear direction. However, a reduced displacement also decreases the denominator in the Eq. (12), which in turn amplifies noise during reconstruction. Therefore, the choice of lateral displacement must balance spatial resolution and robustness to noise. In 2024, Li et al. \cite{li2024single} enhanced this scheme and proposed a complex amplitude imaging architecture based on single metalens, as shown in Fig. 17. By employing a single layer metalens, two sets of shear interference images of a target object along the $x$ and $y$ axes can be simultaneously recorded using a polarization camera, enabled by spatial multiplexing. Each shear interference pattern comprises LCP and RCP components with shear displacements $\Delta_s$ along the $x$ and $y$ directions, respectively. These $x$ and $y$ directional shear patterns are projected onto the four polarization detection channels of the polarization camera, corresponding to 0°, 45°, 90°, and 135° orientations (Fig. 17(a)). Based on the principle of four-wave shear interference, a polarization-dependent PB phase metasurface is used to simultaneously obtain the shear interference pattern of two orthogonal polarization directions in a single measurement. This allows for the derivation of the phase gradient distribution along the $x$/$y$ directions. The minimum detectable phase gradient is 42 mrad/\textmu m, and a relative phase accuracy of 0.0021$\lambda$ is achieved through the complex amplitude image reconstructed by the two-dimensional phase gradient integral, as shown in Fig. 17(b).

\subsection{Phase-Contrast Imaging: Non-Interferometric Extensions and Metasurface 
Integrated Wavefront Sensing}
Overall, advances in metasurface technology are enhancing and transforming traditional wavefront sensing and phase retrieval methods. These innovations extend beyond the optimization of conventional sensing architectures, introducing new approaches to phase-contrast imaging through phase encoding. The following sections provide a brief overview of recent developments in integrated metasurface imaging schemes and their emerging applications. 
\begin{figure}[htbp]
\centering\includegraphics[width=12cm]{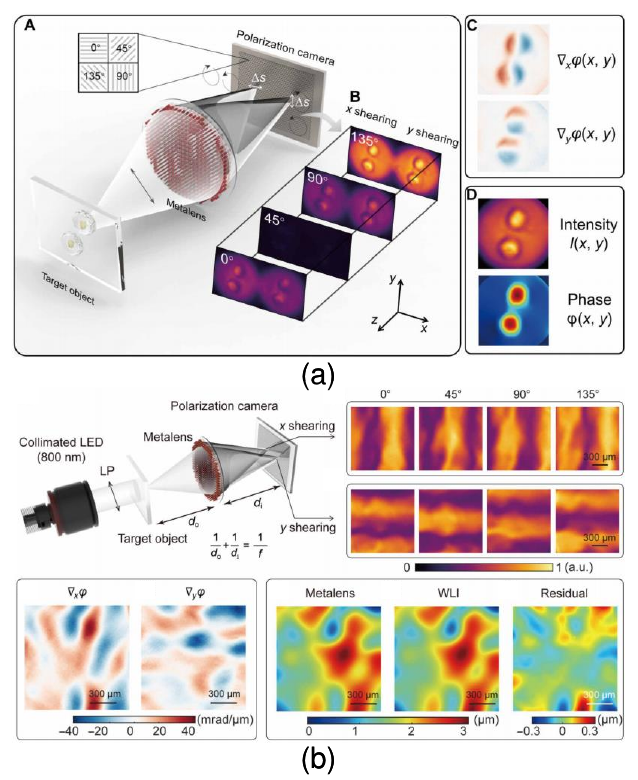}
\caption{Complex amplitude imaging architecture based on a single hyperlens. (a) Operating principle of the hyperlens-assisted single-lens complex amplitude imaging system, including a structural layout, shear interference patterns along the x and y directions at 0°, 45°, 90°, and 135°, and reconstruction of the complex amplitude image from two-dimensional phase gradients (Ref. \cite{li2024single} Fig. 1). (b) Surface characterization using the compact hyperlens-assisted single-lens complex amplitude imaging system, including the system schematic, measured phase gradients along the x and y axes, and surface topography of UV adhesives as measured by the metalens-assisted system, a commercial white-light interferometer, and the corresponding residual maps (Ref. \cite{li2024single} Fig. 3).}
\end{figure}

Phase contrast imaging (PCI) can be viewed as a generalized extension of wavefront sensing techniques within the realm of imaging. Unlike conventional wavefront sensing, which directly quantifies wavefront distortions, PCI enables label-free visualization of transparent specimens—such as biological cells and optical coatings-by converting subtle phase variations into measurable intensity contrast. The underlying principle is that spatially varying phase delays induce coherent interference between diffracted and un-diffracted light components as light traverses a sample with a refractive index gradient. Traditional phase contrast microscopy achieves this by introducing a ring stop and a phase plate at the rear focal plane of the objective lens, applying a fixed phase shift, and attenuating the amplitude of the direct light to convert otherwise invisible phase differences into visible intensity modulations at the image plane. Further details on the technical principles of PCI can be found in the comprehensive review by Marco et al in 2021 \cite{endrizzi2018x}.

In recent years, metasurface technology has revitalized PCI. While conventional systems are constrained by the fixed modulation characteristics of physical phase plates, metasurfaces offer subwavelength-level control over light’s phase, amplitude, and polarization. This enables on-chip integration of functional components-including the ring diaphragm, phase plate, and imaging optics, into a single planar device, enabling ultra-compact, highly versatile PCI systems. In the following sections, we systematically review recent advances in metasurface-enabled PCI and discuss their broader implications for the evolution of wavefront sensing technologies.

Kim et al. overcame the limitations of traditional 4f systems and separately designed spiral phase plates by integrating hyperbolic phase modulation (for beam focusing) and spiral phase modulation (for orbital angular momentum control) onto a single layer metasurface, as illustrated in Figs. 18(a)(b). Kim et al. \cite{kim2022spiral} employed a PB phase metasurface to replicate the target phase distribution. A single layer metasurface was utilized, as illustrated in Fig. 18(c). This approach enabled full-field quantitative phase imaging of biological cells, achieving resolutions of 1.74 \textmu m, 1.23 \textmu m, and 0.78 \textmu m at wavelengths of 497 nm, 532 nm, and 580 nm, respectively, thereby significantly simplifying the detection process for dynamic biological samples. Building upon Kim's research, Xing et al \cite{xingmonolithic}. further realized polarization channel decoupling of optical field modes through spin-multiplexed metalens as shown in Fig. 19(a). The design leverages the independent control capabilities of the PB phase metasurface over left and right circularly polarized light; the electron microscopy image of the metasurface structure is shown in Fig. 19(b). When LCP light is incident, the metasurface applies a hyperbolic phase combined with a constant phase to achieve bright-field imaging. For RCP light, a hyperbolic phase combined with a spiral phase is introduced to generate an edge-enhanced phase contrast image, as illustrated in Figs. 19(c). Experimental results demonstrate that both imaging modes can be obtained simultaneously in a single exposure, eliminating the need for mechanical stretching, phase-change materials, and additional imaging modules. The achieved spatial resolution exceeds 4.4 \textmu m. This multimodal imaging mechanism, enabled by parallel polarization states, offers a hardware-efficient solution for real-time dynamic observation of living biological samples. 
\begin{figure}[htbp]
\centering\includegraphics[width=12cm]{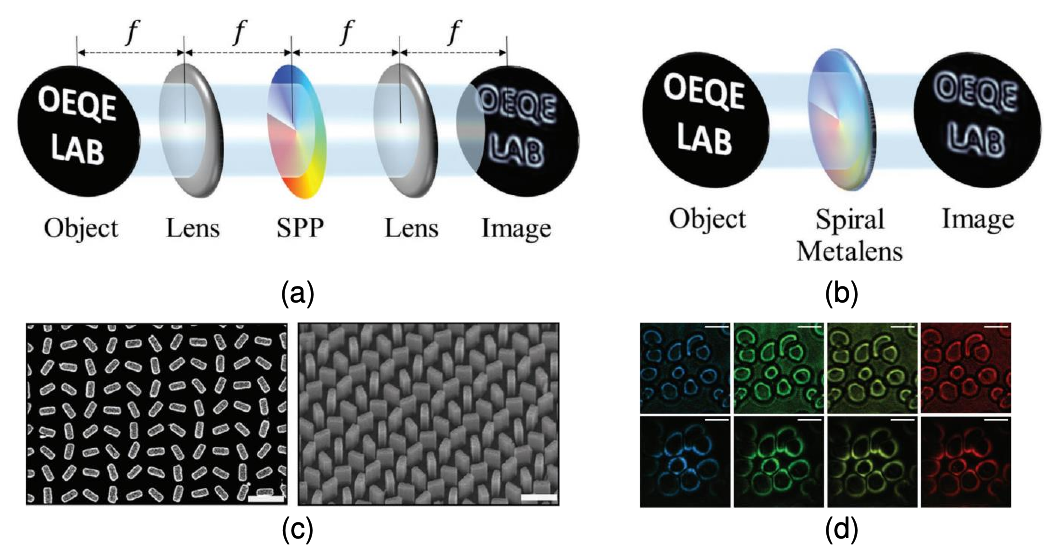}
\caption{PCI scheme based on a metalens. (a) Schematic of a conventional phase-contrast imaging optical path. A 4$f$ optical system is employed, with a phase modulator inserted at the Fourier plane to enable phase-to-intensity conversion at the detector (Ref. \cite{kim2022spiral}, Fig. 1(b)). (b) Schematic of a phase-contrast imaging path based on a metalens, where incident light passes directly through the hyperlens to form a phase-contrast image on the detector (Ref. \cite{kim2022spiral}, Fig. 1(c)). (c) Metalens-assisted electron microscopy (Ref. \cite{kim2022spiral}, Fig. 3(d)). (d) Phase-contrast imaging of biological cells (Ref. \cite{kim2022spiral}, Fig. 5).}
\end{figure}

\begin{figure}[htbp]
\centering\includegraphics[width=12cm]{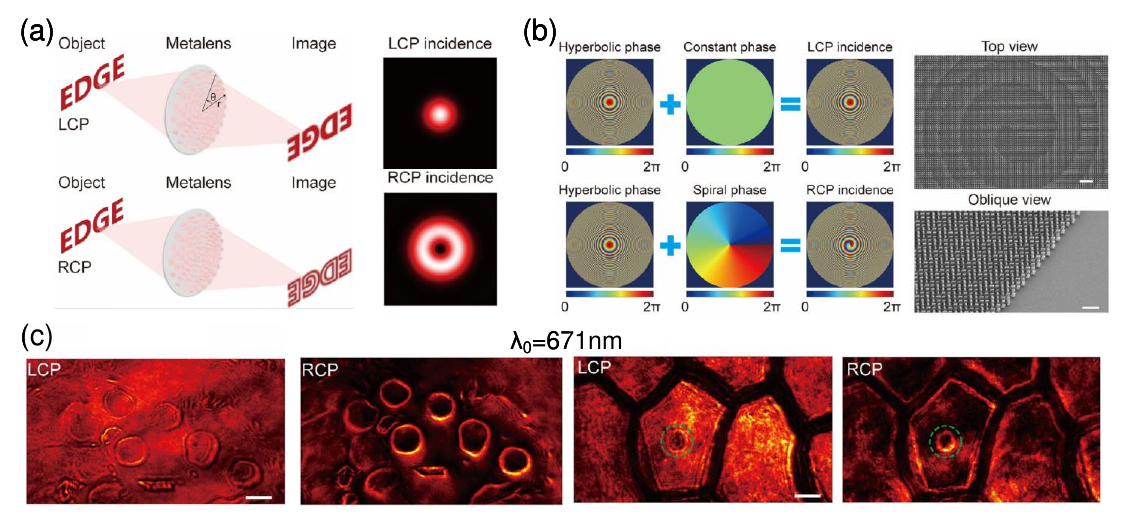}
\caption{Dual-function PCI scheme based on a metalens. (a) Schematic of the dual-function PCI system. LCP light results in bright-field imaging, enabling direct visualization of phase objects, whereas RCP light yields PCI imaging, enhancing image edges (Ref.\cite{xingmonolithic} Fig. 1(a)); (b) Phase profiles of the metalens for LCP and RCP illumination, alongside a scanning electron microscope image of the fabricated metalens structure (Ref.\cite{xingmonolithic} Fig. 1(b), Fig. 2(f)); (c) Imaging results of frog tongue epithelial cells and onion epidermal cells (Ref.\cite{xingmonolithic} Figs. 5(b)(c)).}
\end{figure}

In the optimization of phase imaging systems, Ji et al. \cite{ji2022quantitative} proposed a surface-enhanced 4f phase contrast imaging scheme based on guided mode resonators in 2022, as shown in Fig. 20(a). This study overcomes the limitations of local optical field modulation in traditional Fourier filters by leveraging the nonlocal optical response of a guided mode resonator metasurface (lattice constant: 360 nm, height: 230 nm) to achieve high-contrast phase filtering in the spectral plane of a 4$f$ system, as illustrated in Fig. 20(b). The angular filtering approach employed in this scheme offers three key advantages that are rarely achievable with conventional optical microscopes. First, angular filters introduce minimal absorption or scattering of light. This feature enables dynamic modulation of the effective focal length across different regions of the sample, allowing the realization of various 4$f$ imaging configurations without the need to reposition optical components. Second, phase variations introduced during light propagation can be visualized directly optical manipulation at the Fourier plane is required. Instead, a contrast-enhancing microscope separates background illumination from diffracted light, rendering phase differences observable. Third, the optical alignment process is significantly simplified.  Unlike conventional Fourier filters that demand precise placement, angular filters can be positioned arbitrarily; upon recombination, they inherently generate high-contrast interference, facilitating robust and flexible imaging performance. Experimental results demonstrate that the scheme achieves a phase detection sensitivity of up to 0.02$\pi$, enabling clear differentiation between polystyrene nanospheres and biological cells, as shown in Fig. 20(c). This result highlights the unique advantages of metasurfaces in extending the performance boundaries of traditional imaging systems, particularly for high-precision quantitative phase detection of weakly scattered samples. 

Vortex light, a form of structured light, is characterized by a helical phase distribution and carries a conserved quantity known as the topological charge. Measuring the topological charge and polarization order of vortex beams has long posed a challenge in the field. Indirect measurement techniques- converting phase information into intensity distributions—offer a promising approach for quantifying the topological charge and polarization order of vortex light. Fu et al. \cite{fu2019measuring} introduced a compact optical scheme for measuring phase and polarization singularities based on spin-multiplexed metasurfaces, as shown in Fig. 21(a). When circularly polarized light is incident on the metasurface, the output beam is deflected differently depending on its polarization state, as shown in Fig. 21(b). Due to the phase modulation induced by the metasurface, the left- and right-handed output beams generate distinct far-field optical vortex arrays, each carrying different orbital angular momentum (OAM). If the incident circularly polarized light possesses a helical wavefront with OAM, the vortex array reconfigures into a bright Gaussian point at a specific location determined by the OAM value of the incident beam. By detecting the position of these recovered Gaussian points within the far-field optical vortex array, the topological charge of the vortex beam and the polarization order of the cylindrical vector beam can be individually or simultaneously measured.
\begin{figure}[ht]
\centering\includegraphics[width=12cm]{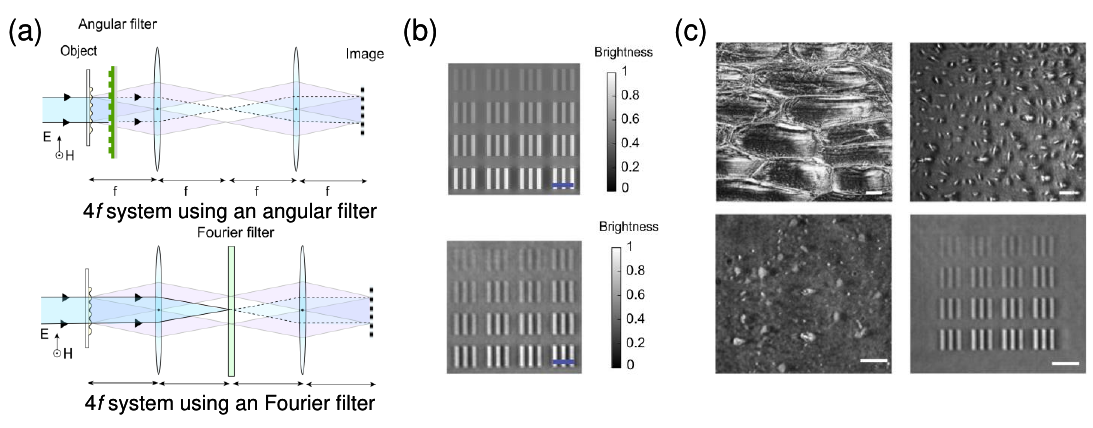}
\caption{PCI scheme employing non-local angular filtering based on a metalens. (a) Top: Schematic of a 4$f$ imaging system incorporating non-local angular filtering via a metalens (Ref. \cite{ji2022quantitative} Fig. 2(a)). Bottom: Traditional 4$f$ imaging system employing conventional Fourier filtering (Ref. \cite{ji2022quantitative} Fig. 2(a)); (b) Top: Simulated and corresponding phase contrast imaging (PCI) results. Bottom: Experimentally acquired PCI images (Ref. \cite{ji2022quantitative} Figs. 5(c)(d)); (c) Phase contrast images of various specimens, including onion epidermal cells (top left), human osteosarcoma cells (top right), hexagonal boron nitride sheets (bottom left), and silicon nitride cylinders fabricated using commercial phase plates with varying diameters, along with their corresponding PCI images (bottom right) (Ref. \cite{ji2022quantitative} Fig. 1(b)).}
\end{figure}

Overall, PCI offers an intuitive and efficient approach to visualizing phase information for wavefront sensing by directly converting phase variations into measurable intensity contrast. Compared to phase retrieval algorithms-which rely on computationally intensive iterative calculations, traditional interferometric techniques that demand precise optical alignment, PCI exhibits several core advantages: 1) Non-interferometric robustness: PCI does not require reference-beam interference, thereby avoiding the stringent demands on optical path stability and coherence. This makes it inherently more robust to environmental perturbations such as vibration and temperature fluctuations; 2) Simplified hardware and reduced cost: Traditional interferometry typically relies on beam splitters, reference mirrors, and other complex components. In contrast, PCI achieves light-field modulation using phase plates or metasurfaces, significantly reducing system complexity; 3) Lightweight computation: Unlike phase retrieval techniques, which require multi-frame iterative optimization, PCI can extract qualitative or semi-quantitative phase information from a single-frame intensity image, enabling real-time wavefront analysis. These features make PCI particularly attractive for applications demanding high temporal resolution and environmental tolerance, such as biological imaging and industrial in-line inspection. More broadly, PCI illustrates a paradigm shift in wavefront sensing from indirect computational reconstruction to direct optical observation-opening new avenues for compact, real-time optical diagnostics.

\begin{figure}[htbp]
\centering\includegraphics[width=12cm]{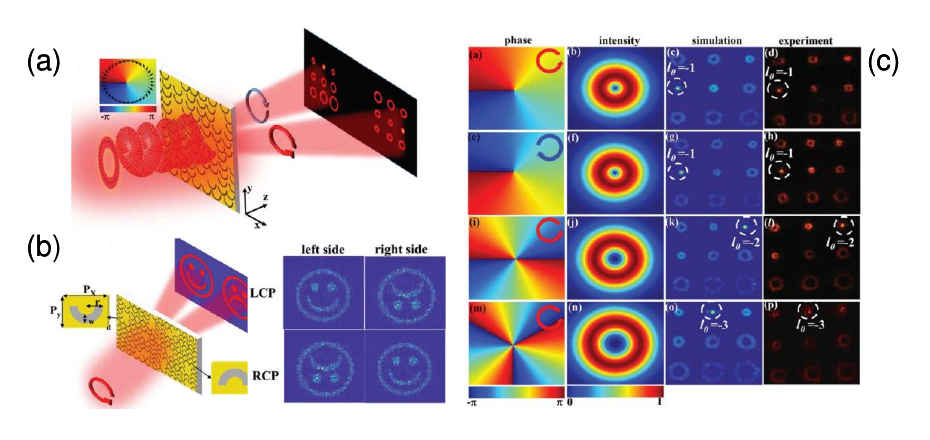}
\caption{A scheme for simultaneously measuring the phase and polarization of vortex beams is proposed based on spin-multiplexed metasurfaces. (a) The principle of this method relies on spin-dependent light–matter interaction within the metasurface structure, enabling simultaneous retrieval of the vortex beam’s phase and polarization state (Ref. \cite{fu2019measuring}, Fig. 1(a)); (b) A schematic illustration of the multiplexing mechanism is shown. Each unit cell of the metasurface consists of two oppositely oriented half-ring apertures. Upon illumination with left-circularly polarized light, distinct far-field patterns—depicted as "happy" and "sad" faces—are reconstructed in separate regions, demonstrating the spatial separation of spin components (Ref. \cite{fu2019measuring}, Fig. 2(a)); (c) FDTD simulations and experimental results confirm the ability to detect phase singularities of vortex beams at a wavelength of 633 nm. The system produces Gaussian-like focal spots whose positions vary with the topological charge of the incident vortex field, validating the scheme’s capability for mode-resolved detection (Ref. \cite{fu2019measuring}, Fig. 4).}
\end{figure}

\subsection{New Concepts in Wavefront Sensing and Phase Retrieval: Multidimensional Control and Quantum State Encoding via Metasurfaces}
The fusion of metasurfaces with traditional wavefront sensing schemes has led to the development of novel wavefront sensors that exhibit both superior performance and compact structures. Moreover, the unique capability of metasurface nanostructures to manipulate light fields has engendered a range of new physical phenomena, offering not only fresh insights into wavefront sensing but also advancing the development of unconventional wavefront detection methods. Building on these advancements, this paper presents several emerging wavefront (phase) retrievals schemes and elucidates their underlying principles. Particular attention is given to approaches that encode quantum states and extract phase information of the light field from quantum systems.

The first method is a wavefront sensing technique based on surface plasmon waves. Plasmonic devices have been utilized to convert electromagnetic waves into free space into bound surface waves, known as surface plasmon waves (SP). Within the subwavelength structure of the experimental surface, the excited surface plasmon wave retains the wavefront information of the incident wave \cite{pelzman2019wavefront}. The excitation of the subwavelength aperture's inner surface plasmon wave can be described by the Huygens-Fresnel principle. Fig. 22(a) illustrates the excitation of surface plasmon waves by a single aperture. According to Huygens' principle, when the incident wave interacts with the subwavelength structure, each point along the aperture can be regarded as a secondary spherical wave, and the SP excited by the concept power supply on the surface can be expressed as:
\begin{eqnarray}
\varphi_{s p p}(x, y, z)=e^{-k} \int_{\phi=0}^{2 \pi}\left(\frac{\varphi_{0}(\rho, \phi) e^{-j\left(\beta_{s p p} \sqrt{(x-\Delta x)^{2}+(y-\Delta y)^{2}}+k_{0} S(\rho, \phi)\right)}}{\sqrt{(x-\Delta x)^{2}+(y-\Delta y)^{2}}}\right) d \phi,
\end{eqnarray}

where, $\beta_{spp}$=2$\pi$/$\lambda_{spp}$ denotes the propagation constant of the surface plasmon wave, while $\varphi_0$($\rho$,$\phi$) represents the amplitude of the light field, determined by the excitation efficiency of the secondary wavelet in generating the surface plasmon wave. $k_0$$S$($\rho$,$\phi$) corresponds to the phase factor of the conceptual point source at the location of the incident wave in ($\rho$,$\phi$), whereas $k_0=2\pi /\lambda_0$ denotes the propagation constant of the incident wave in free space. $k=\sqrt{\beta^{2}{ }_{s p p}-k^{2}{ }_{0} \varepsilon_{c}}$ characterizes the transverse component of the surface plasmon wave vector, which is further described by two parameters: $\Delta x=\rho\cos{\phi}$ and $\Delta t=\rho\sin{\phi}$. 

The shape of the incident wavefront is determined by the constant $S$($\rho$,$\phi$). When the incident wavefront is not perfectly parallel to the $x$-$y$ plane, the secondary wavelets generated by the interaction between the incident wave and the aperture exhibit phase delays due to angular variations. The surface plasmon wave excited by the aperture can be modeled as a two-dimensional circular wave. Multiple surface plasmon waves excited along the aperture interfere with each other, forming a focal point within the circular region. Modifying the shape of the incident wavefront alters the phase distribution of the secondary wavelets along the aperture, resulting in a spatial shift of the focal point within the annular aperture. The displacement $\Delta d$ of the focal point can be determined using the following expression:
\begin{eqnarray}
\Delta d=\Delta \alpha / 2 \beta_{spp} ,
\end{eqnarray}
where $\Delta \alpha$ is phase difference.

In 2019, Pelzman et al. proposed a method for wavefront detection using nanoscale apertures to excite surface plasmon waves and observe the relative displacement of the focused SP waves \cite{pelzman2019wavefront}, as illustrated in Fig. 22(b). Through experimental analysis, they established the relationship between the focal spot displacement ($\Delta d$) and the relative phase difference ($\Delta \alpha$), as shown in Fig. 22(c).

\begin{figure}[ht]
\centering\includegraphics[width=12cm]{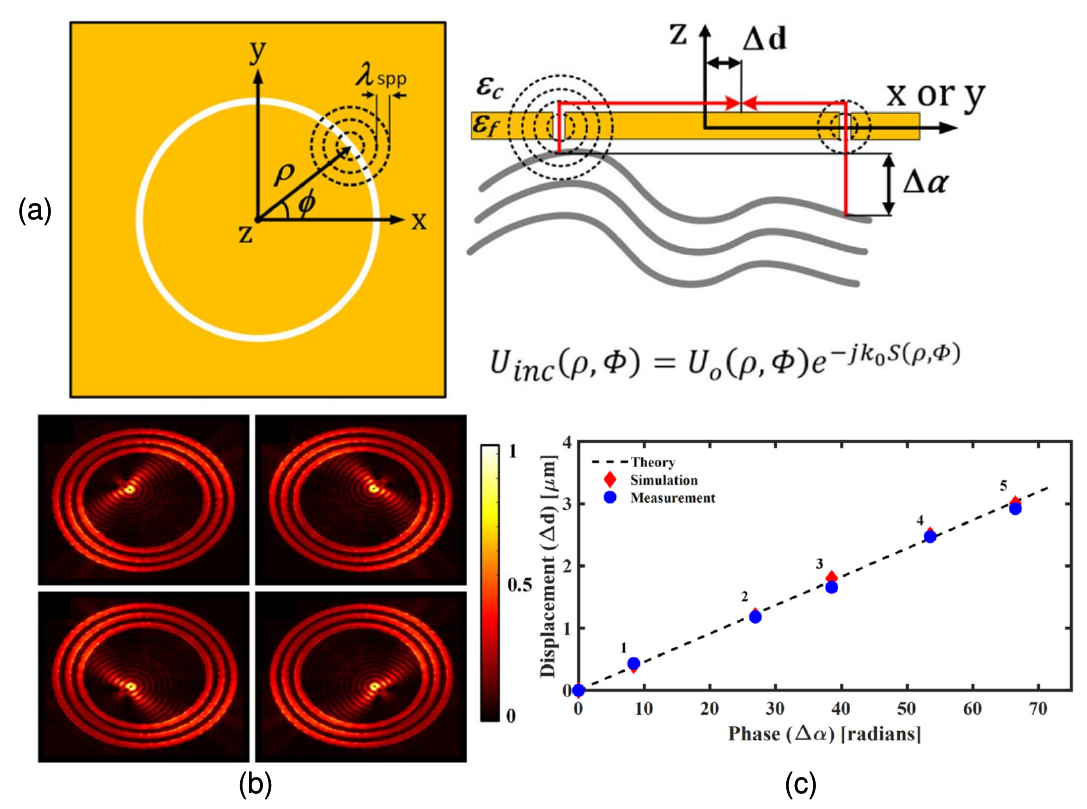}
\caption{Overview of surface plasmon wave formation and wavefront detection principles. (a) In the circular aperture configuration, surface plasmon waves are generated by secondary wavelets originating from the surface boundary, illustrated through both top-view and side-view schematics (Ref. \cite{pelzman2019wavefront} Fig. 1). (b) Displacement of the surface plasmon wave focal point as a function of the incident wavefront's azimuthal and polar angles.(Ref. \cite{pelzman2019wavefront} Fig. 2) (c) Theoretical, simulated, and experimental results of surface plasmon wave displacement and phase shift under a given phase modulation (Ref. \cite{pelzman2019wavefront} Fig. 3(a)).}
\end{figure}

The second approach is based on phase recovery through weak measurement techniques. In quantum mechanics, the act of measuring a wavefunction induces an instantaneous and irreversible collapse of the system, known as wavefunction collapse. The measurement process is represented by an operator, and when the system collapses to an eigenstate of the operator, the original wavefunction information is lost. Weak measurement, however, enables the gradual extraction of information from a quantum system through weak coupling, thereby preventing wavefunction collapse. Since this technique was first introduced by Aharonov et al \cite{aharonov1988result}.It has been widely applied in precision metrology. The weak value is defined as:
\begin{eqnarray}
A_{w}=\frac{\left\langle\psi_{f}\right| \hat{A}\left|\psi_{i}\right\rangle}{\left\langle\psi_{f}\right|\left|\psi_{i}\right\rangle},
\end{eqnarray}
where $\hat{A}$ represents the observed value, $\psi_{f}$ denotes the pre-selected state, and $\psi_{i}$ corresponds to the post-selected state. When the pre-selected and post-selected states are nearly orthogonal, a significant amplification effect occurs, enabling the measurement of various physical quantities by tuning the weak values. 

\begin{figure}[ht]
\centering\includegraphics[width=12cm]{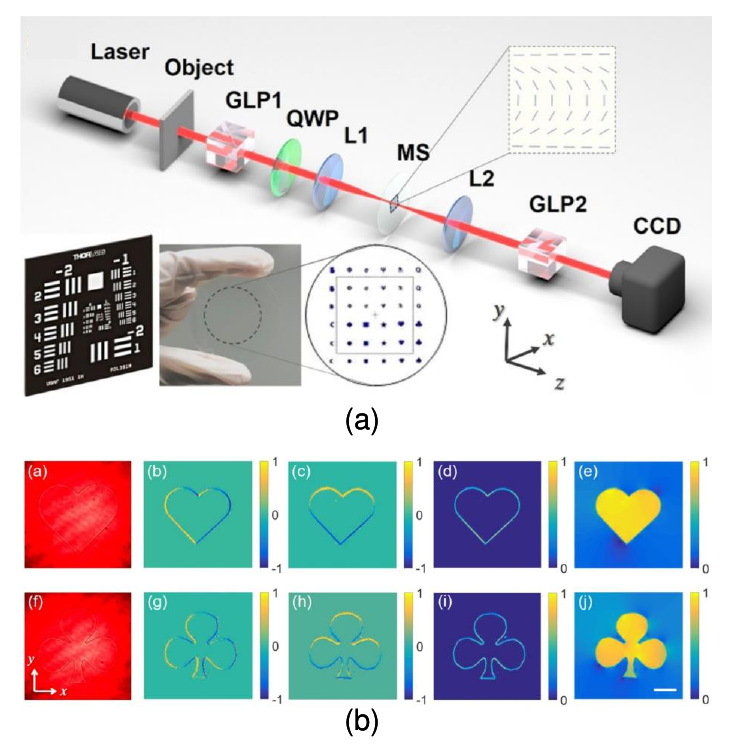}
\caption{Phase reconstruction scheme based on weak measurement and summary of results. (a) Experimental setup. A helium-neon laser generates a linearly polarized Gaussian beam. The combination of GLP1 and QWP enables the beam to be set to any desired elliptical polarization state. Two lenses perform the Fourier and inverse Fourier transforms of the beam. After passing through the system, the output intensity is recorded by a CCD. The inset in the upper right illustrates the spatial distribution of the metasurface’s local optical axis as a function of its periodicity (Ref. \cite{luo2024phase} Fig. 2(b)). (b) Phase reconstruction results. Experimental results for heart-shaped and blade-shaped phase reconstructions, including target phase profiles, first and second derivatives along the x and y directions, and the reconstructed phase distributions (Ref. \cite{luo2024phase} Fig. 5).}
\end{figure}

Luo et al \cite{luo2024phase}. proposed a complex amplitude reconstruction scheme for weak measurements based on a PB phase metasurface in 2024. This metasurface is composed of a uniform material with a phase delay of $\pi$, with its optical axis direction varying periodically along the Y-axis. When a spin-polarized photon passes through the metasurface, it acquires a spatially varying phase. The metasurface introduces weak coupling, which can be easily controlled. By adjusting the pre-selection and post-selection states, both a weak value and a real weak value can be obtained. The phase and amplitude of the optical field are then reconstructed using the weak value and real weak value, respectively. Luo et al. experimentally demonstrated the feasibility of this approach and successfully reconstructed various complex amplitude patterns, as illustrated in Fig. 23.

\begin{figure}[ht]
\centering\includegraphics[width=12cm]{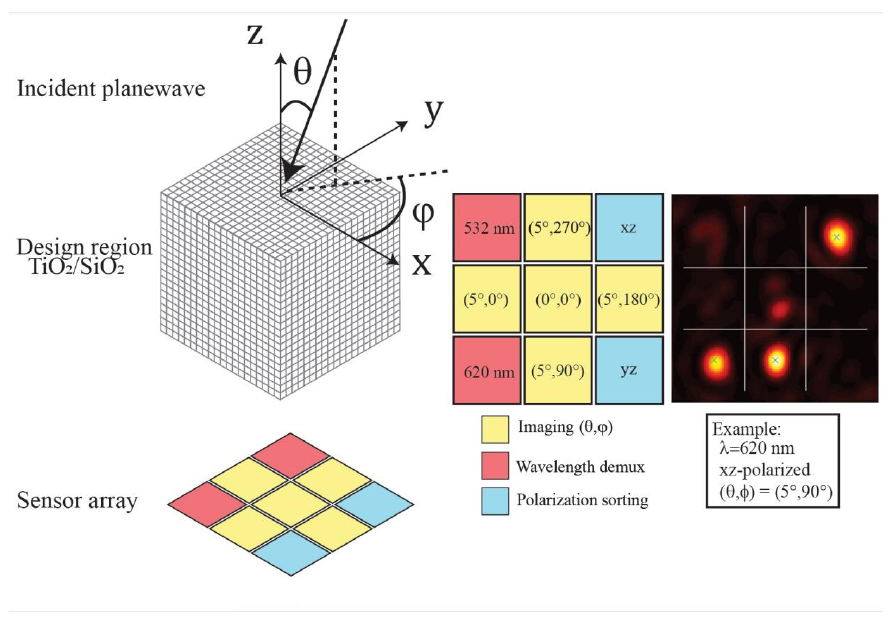}
\caption{Overview of multi-dimensional wavefront sensing driven by volumetric meta-optical elements. This system features a multi-dimensional volumetric meta-optical layout, with the device positioned above a sensor array. An incident plane wave enters at an angle ($\theta$, $\phi$) and interacts with a volumetric meta-optical element composed of silica and titanium dioxide, whose base is located 1.5 \textmu m above the sensor array. The device focuses light onto different sensor pixels depending on the state of the incoming wavefront. For instance, when an incident light wave with a wavelength of 620 nm, an incident angle of (5°, 90°), and polarization in the $xz$ plane enters the volumetric meta-optical element, distinct energy distributions appear at the 3rd, 7th, and 8th positions of the detector. This spatial intensity distribution enables the decoding of information about the incident optical field (Ref. \cite{ballew2023multi} Fig. 1)}
\end{figure}

The multi-dimensional control capability of metasurfaces enables optical systems to modulate multiple degrees of freedom, including propagation direction, wavelength, and polarization—and map them onto a two-dimensional sensor. However, when independent phase modulation is extended from two orthogonal polarization states to multiple angles and wavelengths, it often comes at the cost of efficiency. This trade-off can be mitigated by cascading metasurfaces, though doing so requires time-consuming finite-time-domain simulations of multiple layers. A promising alternative lies in the integration of three-dimensional optical crystals and metaatoms. In 2023, Ballew et al. \cite{ballew2023multi} demonstrated through simulations that a reverse-designed, highly scattering volumetric element composed of subwavelength media can effectively classify light based on direction, wavelength, and polarization. This approach provides a novel framework for wavefront sensing, beam analysis, and next-generation multi-channel optical sensors by leveraging different pixel combinations, as illustrated in Fig. 24. They formulated the design process as an optimization problem by constructing a mapping between plane waves of specific angles, polarizations, and wavelengths and the corresponding detector pixels, expressed in the form of a loss function. Unlike conventional schemes, where pixel response is characterized by intensity at a particular location, their approach defines the optimal value function based on power transfer, which more accurately represents the pixel's response to the incident signal. This function is iteratively optimized using topology optimization methods \cite{jensen2011topology,ji2023recent,dainese2024shape,phan2019high}, ultimately yielding a hyper-volume optical device that maximizes the desired optical response.

\section{Concluding remarks}
This review first examines the theoretical framework of classical wavefront sensing and phase recovery schemes in adaptive optical systems, including Shack-Hartmann sensors, curvature sensors, pyramid sensors, and shear-based wavefront sensors, summarizing research progress over the past decade. Additionally, the innovative applications of various wavefront sensing techniques across different fields are discussed. Next, the integration of metasurface technology with traditional wavefront sensing methods is reviewed. On the one hand, metasurfaces offer a high degree of freedom in light-field modulation, allowing the replacement of multiple conventional lenses within optical systems, thereby significantly reducing system size. Furthermore, the subwavelength structure of metasurfaces ensures precise modulation, and their integration with classical wavefront sensing schemes markedly enhances overall performance. On the other hand, the multi-functionality of metasurfaces has drawn increasing attention to wavefront sensing schemes based on phase retrieval. The core principle of such schemes involves modulating an unknown light field with several known, uncorrelated modes. After capturing multiple-intensity images, the phase of the light field is iteratively reconstructed using regularization constraints and optimization algorithms, ultimately recovering the wavefront at the optical system’s exit pupil. Metasurfaces greatly enhance the efficiency of phase retrieval by enabling the simultaneous acquisition of multiple-intensity distributions through multifunctional modulation.

Beyond the technical optimization of traditional schemes enabled by the exceptional control capabilities of metasurfaces, the modulation of light fields through subwavelength microstructures also gives rise to novel physical mechanisms, such as quantum-state-mediated light-field manipulation. This, in turn, has led to the development of new wavefront sensing paradigms. The enhanced precision in light-field control allows wavefront sensing systems to establish correlations between wavefronts and more sensitive physical quantities, surpassing the resolution limits of conventional optical components while maintaining system compactness and achieving ultra-high spatial resolution. This breakthrough in resolution highlights the transformative potential of metasurfaces in advancing wavefront sensing technologies. Currently, metasurface-based wavefront sensing remains in its infancy. While its high integration and design flexibility have garnered significant research interest, underscoring its academic value, its practical feasibility in optical systems has yet to be demonstrated. This is largely due to the limited design methodologies available for metasurfaces and the stringent precision requirements for both detection and fabrication. Meanwhile, global optical device development is trending toward two extremes: larger-scale systems and ultra-high integrated compact designs. We believe that with advancements in micro-nano fabrication, inverse design algorithms, and adaptive control materials, metasurface-based wavefront sensing technology holds great promise for applications in astronomical observation (e.g., common-phase detection in extremely large telescopes), biological microscopy (e.g., real-time wavefront monitoring of living cells), laser beam shaping (e.g., high-energy laser wavefront correction), and optical communications (e.g. mode-division multiplexing channel reconstruction). This technology is poised to drive the evolution of sensors toward on-chip intelligent sensing, offering essential support for the next generation of high-precision, miniaturized, and multifunctional integrated devices.

This review systematically analyzes the core mechanisms and developmental challenges of traditional wavefront sensing while integrating metasurfaces' innovative light-field modulation capabilities. It aims to offer new research directions for designing and applying metasurface-based wavefront sensors. Additionally, it provides insights into optimizing key optical components and enhancing signal processing algorithms in conventional wavefront sensing schemes. We hope that this review not only inspires researchers working on wavefront sensing, but also serves as a valuable reference for those engaged in the design, fabrication and testing of metasurface components.


\bibliography{sample}

\begin{thebibliography}{100}
\newcommand{\enquote}[1]{``#1''}

\bibitem{hampson2021adaptive}
K.~M. Hampson, R.~Turcotte, D.~T. Miller, \emph{et~al.}, \enquote{Adaptive optics for high-resolution imaging,} {\protect\JournalTitle{Nature Reviews Methods Primers}} \textbf{1}, 68 (2021).

\bibitem{hampson2008adaptive}
K.~M. Hampson, \enquote{Adaptive optics and vision,} {\protect\JournalTitle{Journal of Modern Optics}} \textbf{55}, 3425--3467 (2008).

\bibitem{babcock1953possibility}
H.~W. Babcock, \enquote{The possibility of compensating astronomical seeing,} {\protect\JournalTitle{Publications of the Astronomical Society of the Pacific}} \textbf{65}, 229--236 (1953).

\bibitem{babcock1990adaptive}
H.~W. Babcock, \enquote{Adaptive optics revisited,} {\protect\JournalTitle{Science}} \textbf{249}, 253--257 (1990).

\bibitem{beckers1993adaptive}
J.~M. Beckers, \enquote{Adaptive optics for astronomy-principles, performance, and applications,} {\protect\JournalTitle{In: Annual review of astronomy and astrophysics. Vol. 31 (A94-12726 02-90), p. 13-62.}} \textbf{31}, 13--62 (1993).

\bibitem{ji2012characterization}
N.~Ji, T.~R. Sato, and E.~Betzig, \enquote{Characterization and adaptive optical correction of aberrations during in vivo imaging in the mouse cortex,} {\protect\JournalTitle{Proceedings of the National Academy of Sciences}} \textbf{109}, 22--27 (2012).

\bibitem{liu2019direct}
R.~Liu, Z.~Li, J.~S. Marvin, and D.~Kleinfeld, \enquote{Direct wavefront sensing enables functional imaging of infragranular axons and spines,} {\protect\JournalTitle{Nature methods}} \textbf{16}, 615--618 (2019).

\bibitem{miller2020cellular}
D.~T. Miller and K.~Kurokawa, \enquote{Cellular-scale imaging of transparent retinal structures and processes using adaptive optics optical coherence tomography,} {\protect\JournalTitle{Annual review of vision science}} \textbf{6}, 115--148 (2020).

\bibitem{hofer2011wavefront}
H.~Hofer, N.~Sredar, H.~Queener, \emph{et~al.}, \enquote{Wavefront sensorless adaptive optics ophthalmoscopy in the human eye,} {\protect\JournalTitle{Optics express}} \textbf{19}, 14160--14171 (2011).

\bibitem{kong2024adaptive}
W.~Kong, J.~Huang, Y.~He, and G.~Shi, \enquote{Adaptive optics methods to rat eye properties: Impact of pupil diameter on wavefront detection.} in \emph{Photonics,}  vol.~11 (2024).

\bibitem{yao2023construction}
P.~Yao, R.~Liu, T.~Broggini, \emph{et~al.}, \enquote{Construction and use of an adaptive optics two-photon microscope with direct wavefront sensing,} {\protect\JournalTitle{Nature protocols}} \textbf{18}, 3732--3766 (2023).

\bibitem{booth2007adaptive}
M.~J. Booth, \enquote{Adaptive optics in microscopy,} {\protect\JournalTitle{Philosophical Transactions of the Royal Society A: Mathematical, Physical and Engineering Sciences}} \textbf{365}, 2829--2843 (2007).

\bibitem{salter2019adaptive}
P.~S. Salter and M.~J. Booth, \enquote{Adaptive optics in laser processing,} {\protect\JournalTitle{Light: Science \& Applications}} \textbf{8}, 110 (2019).

\bibitem{bechet2014beam}
C.~B{\'e}chet, A.~Guesalaga, B.~Neichel, \emph{et~al.}, \enquote{Beam shaping for laser-based adaptive optics in astronomy,} {\protect\JournalTitle{Optics express}} \textbf{22}, 12994--13013 (2014).

\bibitem{arain2007adaptive}
M.~A. Arain, V.~Quetschke, J.~Gleason, \emph{et~al.}, \enquote{Adaptive beam shaping by controlled thermal lensing in optical elements,} {\protect\JournalTitle{Applied optics}} \textbf{46}, 2153--2165 (2007).

\bibitem{yang2007adaptive}
P.~Yang, Y.~Liu, W.~Yang, \emph{et~al.}, \enquote{An adaptive laser beam shaping technique based on a genetic algorithm,} {\protect\JournalTitle{Chinese Optics Letters}} \textbf{5}, 497--500 (2007).

\bibitem{nie2021adaptive}
J.~Nie, L.~Tian, H.~Wang, \emph{et~al.}, \enquote{Adaptive beam shaping for enhanced underwater wireless optical communication,} {\protect\JournalTitle{Optics express}} \textbf{29}, 26404--26417 (2021).

\bibitem{he2023vectorial}
C.~He, J.~Antonello, and M.~J. Booth, \enquote{Vectorial adaptive optics,} {\protect\JournalTitle{ELight}} \textbf{3}, 23 (2023).

\bibitem{zhang2022application}
H.~Zhang, L.~Xu, Y.~Guo, \emph{et~al.}, \enquote{Application of adamspgd algorithm to sensor-less adaptive optics in coherent free-space optical communication system,} {\protect\JournalTitle{Optics Express}} \textbf{30}, 7477--7490 (2022).

\bibitem{horst2023tbit}
Y.~Horst, B.~I. Bitachon, L.~Kulmer, \emph{et~al.}, \enquote{Tbit/s line-rate satellite feeder links enabled by coherent modulation and full-adaptive optics,} {\protect\JournalTitle{Light: Science \& Applications}} \textbf{12}, 153 (2023).

\bibitem{wang2022orbital}
J.~Wang, J.~Liu, S.~Li, \emph{et~al.}, \enquote{Orbital angular momentum and beyond in free-space optical communications,} {\protect\JournalTitle{Nanophotonics}} \textbf{11}, 645--680 (2022).

\bibitem{chen2019experimental}
M.~Chen, C.~Liu, D.~Rui, and H.~Xian, \enquote{Experimental results of atmospheric coherent optical communications with adaptive optics,} {\protect\JournalTitle{Optics Communications}} \textbf{434}, 91--96 (2019).

\bibitem{li2017adaptive}
S.~Li and J.~Wang, \enquote{Adaptive free-space optical communications through turbulence using self-healing bessel beams,} {\protect\JournalTitle{Scientific reports}} \textbf{7}, 43233 (2017).

\bibitem{extance2015laser}
A.~Extance, \enquote{Laser weapons get real: long a staple of science fiction, laser weapons are edging closer to the battlefield--thanks to optical fibres,} {\protect\JournalTitle{Nature}} \textbf{521}, 408--411 (2015).

\bibitem{scheers2023numerical}
L.~Scheers, G.~Vissers, P.~Piscaer, \emph{et~al.}, \enquote{Numerical analysis of propagation conditions in which conventional adaptive optics can improve laser-directed energy weapons beam shaping,} in \emph{High Power Lasers: Technology and Systems, Platforms, Effects VI,}  (SPIE, 2023), p. PC1273902.

\bibitem{zepp2013holographic}
A.~Zepp, S.~G{\l}adysz, and K.~Stein, \enquote{Holographic wavefront sensor for fast defocus measurement,} {\protect\JournalTitle{Advanced Optical Technologies}} \textbf{2}, 433--437 (2013).

\bibitem{paterson2000hybrid}
C.~Paterson and J.~Dainty, \enquote{Hybrid curvature and gradient wave-front sensor,} {\protect\JournalTitle{Optics letters}} \textbf{25}, 1687--1689 (2000).

\bibitem{hickson1994single}
P.~Hickson and G.~S. Burley, \enquote{Single-image wavefront curvature sensing,} in \emph{Adaptive Optics in Astronomy,}  vol. 2201 (SPIE, 1994), pp. 549--554.

\bibitem{cagigal2015x}
M.~P. Cagigal and P.~J. Valle, \enquote{x--y curvature wavefront sensor,} {\protect\JournalTitle{Optics Letters}} \textbf{40}, 1655--1658 (2015).

\bibitem{liu2025high}
Q.~Liu, M.~Zhang, J.~Tang, \emph{et~al.}, \enquote{High-bandwidth nonlinear curvature wavefront sensing based on dual-defocused scheme and optimal parameter selection,} {\protect\JournalTitle{Optics and Lasers in Engineering}} \textbf{190}, 108975 (2025).

\bibitem{diaz2006curvature}
F.~D{\'\i}az-Dout{\'o}n, J.~Pujol, M.~Arjona, and S.~O. Luque, \enquote{Curvature sensor for ocular wavefront measurement,} {\protect\JournalTitle{Optics letters}} \textbf{31}, 2245--2247 (2006).

\bibitem{letchev2023assessing}
S.~Letchev, J.~Crass, and J.~R. Crepp, \enquote{Assessing phase reconstruction accuracy for different nonlinear curvature wavefront sensor configurations,} {\protect\JournalTitle{Journal of Astronomical Telescopes, Instruments, and Systems}} \textbf{9}, 049001--049001 (2023).

\bibitem{goloborodko2023wavefront}
A.~Goloborodko, \enquote{Wavefront curvature restoration by a sensor based on the talbot phenomenon under gaussian illumination,} {\protect\JournalTitle{Journal of the Optical Society of America A}} \textbf{40}, B8--B14 (2023).

\bibitem{zhong2022hybrid}
H.~Zhong, Y.~Li, P.~Qin, \emph{et~al.}, \enquote{Hybrid wavefront reconstruction from multi-directional slope and full curvature measurements using integral equations with higher-order truncation errors for wavefront sensors,} {\protect\JournalTitle{Optics and Lasers in Engineering}} \textbf{154}, 106991 (2022).

\bibitem{frazin2018efficient}
R.~A. Frazin, \enquote{Efficient, nonlinear phase estimation with the nonmodulated pyramid wavefront sensor,} {\protect\JournalTitle{Journal of the Optical Society of America A}} \textbf{35}, 594--607 (2018).

\bibitem{guthery2021pyramid}
C.~E. Guthery and M.~Hart, \enquote{Pyramid and shack--hartmann hybrid wave-front sensor,} {\protect\JournalTitle{Optics Letters}} \textbf{46}, 1045--1048 (2021).

\bibitem{agapito2023non}
G.~Agapito, E.~Pinna, S.~Esposito, \emph{et~al.}, \enquote{Non-modulated pyramid wavefront sensor-use in sensing and correcting atmospheric turbulence,} {\protect\JournalTitle{Astronomy \& Astrophysics}} \textbf{677}, A168 (2023).

\bibitem{guzman2024deep}
F.~Guzm{\'a}n, J.~Tapia, C.~Weinberger, \emph{et~al.}, \enquote{Deep optics preconditioner for modulation-free pyramid wavefront sensing,} {\protect\JournalTitle{Photonics Research}} \textbf{12}, 301--312 (2024).

\bibitem{bond2020adaptive}
C.~Z. Bond, S.~Cetre, S.~Lilley, \emph{et~al.}, \enquote{Adaptive optics with an infrared pyramid wavefront sensor at keck,} {\protect\JournalTitle{Journal of Astronomical Telescopes, Instruments, and Systems}} \textbf{6}, 039003--039003 (2020).

\bibitem{bertrou2022confusion}
A.~Bertrou-Cantou, E.~Gendron, G.~Rousset, \emph{et~al.}, \enquote{Confusion in differential piston measurement with the pyramid wavefront sensor,} {\protect\JournalTitle{Astronomy \& Astrophysics}} \textbf{658}, A49 (2022).

\bibitem{chambouleyron2020pyramid}
V.~Chambouleyron, O.~Fauvarque, P.~Janin-Potiron, \emph{et~al.}, \enquote{Pyramid wavefront sensor optical gains compensation using a convolutional model,} {\protect\JournalTitle{Astronomy \& Astrophysics}} \textbf{644}, A6 (2020).

\bibitem{lozi2019visible}
J.~Lozi, N.~Jovanovic, O.~Guyon, \emph{et~al.}, \enquote{Visible and near-infrared laboratory demonstration of a simplified pyramid wavefront sensor,} {\protect\JournalTitle{Publications of the Astronomical Society of the Pacific}} \textbf{131}, 044503 (2019).

\bibitem{akondi2013digital}
V.~Akondi, S.~Castillo, and B.~Vohnsen, \enquote{Digital pyramid wavefront sensor with tunable modulation,} {\protect\JournalTitle{Optics express}} \textbf{21}, 18261--18272 (2013).

\bibitem{ragazzoni2002pyramid}
R.~Ragazzoni, E.~Diolaiti, and E.~Vernet, \enquote{A pyramid wavefront sensor with no dynamic modulation,} {\protect\JournalTitle{Optics communications}} \textbf{208}, 51--60 (2002).

\bibitem{meimon2014sensing}
S.~Meimon, T.~Fusco, V.~Michau, and C.~Plantet, \enquote{Sensing more modes with fewer sub-apertures: the lifted shack--hartmann wavefront sensor,} {\protect\JournalTitle{Optics Letters}} \textbf{39}, 2835--2837 (2014).

\bibitem{aftab2018adaptive}
M.~Aftab, H.~Choi, R.~Liang, and D.~W. Kim, \enquote{Adaptive shack-hartmann wavefront sensor accommodating large wavefront variations,} {\protect\JournalTitle{Optics Express}} \textbf{26}, 34428--34441 (2018).

\bibitem{basavaraju2014myopic}
R.~M. Basavaraju, V.~Akondi, S.~J. Weddell, and R.~P. Budihal, \enquote{Myopic aberrations: Simulation based comparison of curvature and hartmann shack wavefront sensors,} {\protect\JournalTitle{Optics Communications}} \textbf{312}, 23--30 (2014).

\bibitem{deng2025measurement}
Z.~Deng, C.~Li, and S.~Zhang, \enquote{Measurement of lens parameters based on shack-hartmann wavefront sensor,} {\protect\JournalTitle{Optics and Lasers in Engineering}} \textbf{184}, 108666 (2025).

\bibitem{he2024accuracy}
Y.~He, M.~Bao, Y.~Chen, \emph{et~al.}, \enquote{Accuracy characterization of shack--hartmann sensor with residual error removal in spherical wavefront calibration,} {\protect\JournalTitle{Light: Advanced Manufacturing}} \textbf{4}, 393--403 (2024).

\bibitem{li2021piston}
X.~Li, X.~Yang, S.~Wang, \emph{et~al.}, \enquote{The piston error recognition technique used in the modified shack--hartmann sensor,} {\protect\JournalTitle{Optics Communications}} \textbf{501}, 127388 (2021).

\bibitem{zhang2024automatic}
Q.~Zhang, H.~Zuo, X.~Cui, \emph{et~al.}, \enquote{Automatic compressive sensing of shack--hartmann sensors based on the vision transformer,} in \emph{Photonics,}  vol.~11 (MDPI, 2024), p. 998.

\bibitem{yang2022method}
W.~Yang, J.~Wang, and B.~Wang, \enquote{A method used to improve the dynamic range of shack--hartmann wavefront sensor in presence of large aberration,} {\protect\JournalTitle{Sensors}} \textbf{22}, 7120 (2022).

\bibitem{xiya2023research}
W.~Xiya, S.~Qilin, Y.~Jinsheng, \emph{et~al.}, \enquote{Research on wavefront measurement technology of space-based telescope using shack-hartmann wavefront sensor,} {\protect\JournalTitle{Opto-Electronic Engineering}} \textbf{50}, 230215--1 (2023).

\bibitem{hartlieb2024large}
S.~Hartlieb, Z.~Wang, A.~R{\"u}dinger, \emph{et~al.}, \enquote{Large dynamic range shack--hartmann wavefront sensor based on holographic multipoint generation and pattern correlation,} {\protect\JournalTitle{Optical Engineering}} \textbf{63}, 024107--024107 (2024).

\bibitem{wu2023study}
X.~Wu, L.~Huang, N.~Gu, \emph{et~al.}, \enquote{Study of a shack-hartmann wavefront sensor with adjustable spatial sampling based on spherical reference wave,} {\protect\JournalTitle{Optics and Lasers in Engineering}} \textbf{160}, 107289 (2023).

\bibitem{deng2025sequential}
Y.~Deng, W.~He, L.~Xin, \emph{et~al.}, \enquote{Sequential spot alignment algorithm: Enhancing dynamic range of shack-hartmann sensors,} {\protect\JournalTitle{Optics and Lasers in Engineering}} \textbf{190}, 108952 (2025).

\bibitem{zheng2021detecting}
Y.~Zheng, M.~Yang, Z.-H. Liu, \emph{et~al.}, \enquote{Detecting momentum weak value: Shack--hartmann versus a weak measurement wavefront sensor,} {\protect\JournalTitle{Optics Letters}} \textbf{46}, 5352--5355 (2021).

\bibitem{wang2021deep}
K.~Wang, M.~Zhang, J.~Tang, \emph{et~al.}, \enquote{Deep learning wavefront sensing and aberration correction in atmospheric turbulence,} {\protect\JournalTitle{PhotoniX}} \textbf{2}, 8 (2021).

\bibitem{wang2015direct}
K.~Wang, W.~Sun, C.~T. Richie, \emph{et~al.}, \enquote{Direct wavefront sensing for high-resolution in vivo imaging in scattering tissue,} {\protect\JournalTitle{Nature communications}} \textbf{6}, 7276 (2015).

\bibitem{yi2021angle}
S.~Yi, J.~Xiang, M.~Zhou, \emph{et~al.}, \enquote{Angle-based wavefront sensing enabled by the near fields of flat optics,} {\protect\JournalTitle{Nature communications}} \textbf{12}, 6002 (2021).

\bibitem{berto2017wavefront}
P.~Berto, H.~Rigneault, and M.~Guillon, \enquote{Wavefront sensing with a thin diffuser,} {\protect\JournalTitle{Optics letters}} \textbf{42}, 5117--5120 (2017).

\bibitem{wu2019wish}
Y.~Wu, M.~K. Sharma, and A.~Veeraraghavan, \enquote{Wish: wavefront imaging sensor with high resolution,} {\protect\JournalTitle{Light: Science \& Applications}} \textbf{8}, 44 (2019).

\bibitem{norris2020all}
B.~R. Norris, J.~Wei, C.~H. Betters, \emph{et~al.}, \enquote{An all-photonic focal-plane wavefront sensor,} {\protect\JournalTitle{Nature Communications}} \textbf{11}, 5335 (2020).

\bibitem{guo2024direct}
Y.~Guo, Y.~Hao, S.~Wan, \emph{et~al.}, \enquote{Direct observation of atmospheric turbulence with a video-rate wide-field wavefront sensor,} {\protect\JournalTitle{Nature Photonics}} \textbf{18}, 935--943 (2024).

\bibitem{li2025orthogonal}
Y.~Li, R.~Zhu, S.~Sui, \emph{et~al.}, \enquote{Orthogonal-based reconfigurable light-controlled metasurface for multichannel amplitude-modulation communication,} {\protect\JournalTitle{Laser \& Photonics Reviews}} \textbf{19}, 2401470 (2025).

\bibitem{kamali2018review}
S.~M. Kamali, E.~Arbabi, A.~Arbabi, and A.~Faraon, \enquote{A review of dielectric optical metasurfaces for wavefront control,} {\protect\JournalTitle{Nanophotonics}} \textbf{7}, 1041--1068 (2018).

\bibitem{zhang2020polarization}
X.~G. Zhang, Q.~Yu, W.~X. Jiang, \emph{et~al.}, \enquote{Polarization-controlled dual-programmable metasurfaces,} {\protect\JournalTitle{Advanced science}} \textbf{7}, 1903382 (2020).

\bibitem{balthasar2017metasurface}
J.~Balthasar~Mueller, N.~A. Rubin, R.~C. Devlin, \emph{et~al.}, \enquote{Metasurface polarization optics: independent phase control of arbitrary orthogonal states of polarization,} {\protect\JournalTitle{Physical review letters}} \textbf{118}, 113901 (2017).

\bibitem{dorrah2021metasurface}
A.~H. Dorrah, N.~A. Rubin, A.~Zaidi, \emph{et~al.}, \enquote{Metasurface optics for on-demand polarization transformations along the optical path,} {\protect\JournalTitle{Nature Photonics}} \textbf{15}, 287--296 (2021).

\bibitem{yu2014flat}
N.~Yu and F.~Capasso, \enquote{Flat optics with designer metasurfaces,} {\protect\JournalTitle{Nature materials}} \textbf{13}, 139--150 (2014).

\bibitem{huang2013three}
L.~Huang, X.~Chen, H.~M{\"u}hlenbernd, \emph{et~al.}, \enquote{Three-dimensional optical holography using a plasmonic metasurface,} {\protect\JournalTitle{Nature communications}} \textbf{4}, 2808 (2013).

\bibitem{zhao2020recent}
R.~Zhao, L.~Huang, and Y.~Wang, \enquote{Recent advances in multi-dimensional metasurfaces holographic technologies,} {\protect\JournalTitle{PhotoniX}} \textbf{1}, 1--24 (2020).

\bibitem{zheng2015metasurface}
G.~Zheng, H.~M{\"u}hlenbernd, M.~Kenney, \emph{et~al.}, \enquote{Metasurface holograms reaching 80\% efficiency,} {\protect\JournalTitle{Nature nanotechnology}} \textbf{10}, 308--312 (2015).

\bibitem{deng2020malus}
L.~Deng, J.~Deng, Z.~Guan, \emph{et~al.}, \enquote{Malus-metasurface-assisted polarization multiplexing,} {\protect\JournalTitle{Light: Science \& Applications}} \textbf{9}, 101 (2020).

\bibitem{yue2022versatile}
Z.~Yue, J.~Li, J.~Liu, \emph{et~al.}, \enquote{Versatile polarization conversion and wavefront shaping based on fully phase-modulated metasurface with complex amplitude modulation,} {\protect\JournalTitle{Advanced Optical Materials}} \textbf{10}, 2200733 (2022).

\bibitem{wang2021resonant}
B.~Wang, K.~Wang, X.~Hong, \emph{et~al.}, \enquote{Resonant nonlinear synthetic metasurface with combined phase and amplitude modulations,} {\protect\JournalTitle{Laser \& Photonics Reviews}} \textbf{15}, 2100031 (2021).

\bibitem{hao2021full}
Z.~Hao, W.~Liu, Z.~Li, \emph{et~al.}, \enquote{Full complex-amplitude modulation of second harmonic generation with nonlinear metasurfaces,} {\protect\JournalTitle{Laser \& Photonics Reviews}} \textbf{15}, 2100207 (2021).

\bibitem{high2015visible}
A.~A. High, R.~C. Devlin, A.~Dibos, \emph{et~al.}, \enquote{Visible-frequency hyperbolic metasurface,} {\protect\JournalTitle{Nature}} \textbf{522}, 192--196 (2015).

\bibitem{park2025tape}
Y.~Park, J.~Kim, Y.~Yang, \emph{et~al.}, \enquote{Tape-assisted residual layer-free one-step nanoimprinting of high-index hybrid polymer for optical loss-suppressed metasurfaces,} {\protect\JournalTitle{Advanced Science}} p. 2409371 (2025).

\bibitem{einck2021scalable}
V.~J. Einck, M.~Torfeh, A.~McClung, \emph{et~al.}, \enquote{Scalable nanoimprint lithography process for manufacturing visible metasurfaces composed of high aspect ratio tio2 meta-atoms,} {\protect\JournalTitle{ACS Photonics}} \textbf{8}, 2400--2409 (2021).

\bibitem{chen2015large}
W.~Chen, M.~Tymchenko, P.~Gopalan, \emph{et~al.}, \enquote{Large-area nanoimprinted colloidal au nanocrystal-based nanoantennas for ultrathin polarizing plasmonic metasurfaces,} {\protect\JournalTitle{Nano letters}} \textbf{15}, 5254--5260 (2015).

\bibitem{makarov2017multifold}
S.~V. Makarov, V.~Milichko, E.~V. Ushakova, \emph{et~al.}, \enquote{Multifold emission enhancement in nanoimprinted hybrid perovskite metasurfaces,} {\protect\JournalTitle{ACS Photonics}} \textbf{4}, 728--735 (2017).

\bibitem{choi2023realization}
H.~Choi, J.~Kim, W.~Kim, \emph{et~al.}, \enquote{Realization of high aspect ratio metalenses by facile nanoimprint lithography using water-soluble stamps,} {\protect\JournalTitle{PhotoniX}} \textbf{4}, 18 (2023).

\bibitem{chen2022large}
C.~Chen, Y.~Liu, Z.-y. Jiang, \emph{et~al.}, \enquote{Large-area long-wave infrared broadband all-dielectric metasurface absorber based on maskless laser direct writing lithography,} {\protect\JournalTitle{Optics Express}} \textbf{30}, 13391--13403 (2022).

\bibitem{hu2021high}
Y.~Hu, L.~Li, R.~Wang, \emph{et~al.}, \enquote{High-speed parallel plasmonic direct-writing nanolithography using metasurface-based plasmonic lens,} {\protect\JournalTitle{Engineering}} \textbf{7}, 1623--1630 (2021).

\bibitem{nivas2025femtosecond}
J.~J. Nivas, G.~P. Papari, M.~Hu, \emph{et~al.}, \enquote{Femtosecond laser direct writing of complementary thz metasurfaces using a structured vortex beam,} {\protect\JournalTitle{Optics \& Laser Technology}} \textbf{181}, 111831 (2025).

\bibitem{sun2025generating}
X.~Sun, K.~Zheng, K.~Liu, and Y.~Zeng, \enquote{Generating vector optical field arrays for laser direct writing chiral nanostructures based on metasurface,} {\protect\JournalTitle{Optics and Lasers in Engineering}} \textbf{186}, 108843 (2025).

\bibitem{zhang2025recent}
B.~Zhang, W.~Yan, and F.~Chen, \enquote{Recent advances in femtosecond laser direct writing of three-dimensional periodic photonic structures in transparent materials,} {\protect\JournalTitle{Advanced Photonics}} \textbf{7}, 034002--034002 (2025).

\bibitem{sin2023high}
Y.~Sin~Tan, H.~Wang, H.~Wang, \emph{et~al.}, \enquote{High-throughput fabrication of large-scale metasurfaces using electron-beam lithography with su-8 gratings for multilevel security printing,} {\protect\JournalTitle{Photonics Research}} \textbf{11}, B103--B110 (2023).

\bibitem{baracu2021silicon}
A.~M. Baracu, M.~A. Avram, C.~Breazu, \emph{et~al.}, \enquote{Silicon metalens fabrication from electron beam to uv-nanoimprint lithography,} {\protect\JournalTitle{Nanomaterials}} \textbf{11}, 2329 (2021).

\bibitem{cao2022tuning}
X.~Cao, Y.~Xiao, Q.~Dong, \emph{et~al.}, \enquote{Tuning metasurface dimensions by soft nanoimprint lithography and reactive ion etching,} {\protect\JournalTitle{Advanced Photonics Research}} \textbf{3}, 2200127 (2022).

\bibitem{briere2019etching}
G.~Bri{\`e}re, P.~Ni, S.~H{\'e}ron, \emph{et~al.}, \enquote{An etching-free approach toward large-scale light-emitting metasurfaces,} {\protect\JournalTitle{Advanced Optical Materials}} \textbf{7}, 1801271 (2019).

\bibitem{baracu2021metasurface}
A.~M. Baracu, C.~A. Dirdal, A.~M. Avram, \emph{et~al.}, \enquote{Metasurface fabrication by cryogenic and bosch deep reactive ion etching,} {\protect\JournalTitle{Micromachines}} \textbf{12}, 501 (2021).

\bibitem{he2020metasurfaces}
J.~He, T.~Dong, B.~Chi, and Y.~Zhang, \enquote{Metasurfaces for terahertz wavefront modulation: a review,} {\protect\JournalTitle{Journal of Infrared, Millimeter, and Terahertz Waves}} \textbf{41}, 607--631 (2020).

\bibitem{wei2020optical}
Q.~Wei, L.~Huang, T.~Zentgraf, and Y.~Wang, \enquote{Optical wavefront shaping based on functional metasurfaces,} {\protect\JournalTitle{Nanophotonics}} \textbf{9}, 987--1002 (2020).

\bibitem{zhao2018continuously}
S.-D. Zhao, A.-L. Chen, Y.-S. Wang, and C.~Zhang, \enquote{Continuously tunable acoustic metasurface for transmitted wavefront modulation,} {\protect\JournalTitle{Physical Review Applied}} \textbf{10}, 054066 (2018).

\bibitem{huang2024microcavity}
S.-H. Huang, H.-P. Su, C.-Y. Chen, \emph{et~al.}, \enquote{Microcavity-assisted multi-resonant metasurfaces enabling versatile wavefront engineering,} {\protect\JournalTitle{Nature Communications}} \textbf{15}, 9658 (2024).

\bibitem{wu2024ultrathin}
X.~Y. Wu, H.~Y. Feng, F.~Wan, \emph{et~al.}, \enquote{An ultrathin, fast-response, large-scale liquid-crystal-facilitated multi-functional reconfigurable metasurface for comprehensive wavefront modulation,} {\protect\JournalTitle{Advanced Materials}} \textbf{36}, 2402170 (2024).

\bibitem{barton2021wavefront}
D.~Barton, M.~Lawrence, and J.~Dionne, \enquote{Wavefront shaping and modulation with resonant electro-optic phase gradient metasurfaces,} {\protect\JournalTitle{Applied Physics Letters}} \textbf{118} (2021).

\bibitem{bomzon2001pancharatnam}
Z.~Bomzon, V.~Kleiner, and E.~Hasman, \enquote{Pancharatnam--berry phase in space-variant polarization-state manipulations with subwavelength gratings,} {\protect\JournalTitle{Optics letters}} \textbf{26}, 1424--1426 (2001).

\bibitem{mehmood2016visible}
M.~Mehmood, S.~Mei, S.~Hussain, \emph{et~al.}, \enquote{Visible-frequency metasurface for structuring and spatially multiplexing optical vortices,} {\protect\JournalTitle{Adv. Mater}} \textbf{28}, 2533--2539 (2016).

\bibitem{khorasaninejad2015broadband}
M.~Khorasaninejad and F.~Capasso, \enquote{Broadband multifunctional efficient meta-gratings based on dielectric waveguide phase shifters,} {\protect\JournalTitle{Nano letters}} \textbf{15}, 6709--6715 (2015).

\bibitem{khorasaninejad2016polarization}
M.~Khorasaninejad, A.~Y. Zhu, C.~Roques-Carmes, \emph{et~al.}, \enquote{Polarization-insensitive metalenses at visible wavelengths,} {\protect\JournalTitle{Nano letters}} \textbf{16}, 7229--7234 (2016).

\bibitem{aieta2015multiwavelength}
F.~Aieta, M.~A. Kats, P.~Genevet, and F.~Capasso, \enquote{Multiwavelength achromatic metasurfaces by dispersive phase compensation,} {\protect\JournalTitle{Science}} \textbf{347}, 1342--1345 (2015).

\bibitem{kang2012wave}
M.~Kang, T.~Feng, H.-T. Wang, and J.~Li, \enquote{Wave front engineering from an array of thin aperture antennas,} {\protect\JournalTitle{Optics express}} \textbf{20}, 15882--15890 (2012).

\bibitem{chen2018huygens}
M.~Chen, M.~Kim, A.~M. Wong, and G.~V. Eleftheriades, \enquote{Huygens’ metasurfaces from microwaves to optics: a review,} {\protect\JournalTitle{Nanophotonics}} \textbf{7}, 1207--1231 (2018).

\bibitem{wong2018perfect}
A.~M. Wong and G.~V. Eleftheriades, \enquote{Perfect anomalous reflection with a bipartite huygens’ metasurface,} {\protect\JournalTitle{Physical Review X}} \textbf{8}, 011036 (2018).

\bibitem{leitis2020all}
A.~Leitis, A.~He{\ss}ler, S.~Wahl, \emph{et~al.}, \enquote{All-dielectric programmable huygens' metasurfaces,} {\protect\JournalTitle{Advanced Functional Materials}} \textbf{30}, 1910259 (2020).

\bibitem{fan2018phototunable}
K.~Fan, J.~Zhang, X.~Liu, \emph{et~al.}, \enquote{Phototunable dielectric huygens' metasurfaces,} {\protect\JournalTitle{Advanced Materials}} \textbf{30}, 1800278 (2018).

\bibitem{gigli2021fundamental}
C.~Gigli, Q.~Li, P.~Chavel, \emph{et~al.}, \enquote{Fundamental limitations of huygens’ metasurfaces for optical beam shaping,} {\protect\JournalTitle{Laser \& Photonics Reviews}} \textbf{15}, 2000448 (2021).

\bibitem{eleftheriades2022prospects}
G.~V. Eleftheriades, M.~Kim, V.~G. Ataloglou, and A.~H. Dorrah, \enquote{Prospects of huygens’ metasurfaces for antenna applications,} {\protect\JournalTitle{Engineering}} \textbf{11}, 21--26 (2022).

\bibitem{jia2015independent}
S.~L. Jia, X.~Wan, D.~Bao, \emph{et~al.}, \enquote{Independent controls of orthogonally polarized transmitted waves using a huygens metasurface,} {\protect\JournalTitle{Laser \& Photonics Reviews}} \textbf{9}, 545--553 (2015).

\bibitem{howes2018dynamic}
A.~Howes, W.~Wang, I.~Kravchenko, and J.~Valentine, \enquote{Dynamic transmission control based on all-dielectric huygens metasurfaces,} {\protect\JournalTitle{Optica}} \textbf{5}, 787--792 (2018).

\bibitem{londono2018broadband}
M.~Londo{\~n}o, A.~Sayanskiy, J.~Araque-Quijano, \emph{et~al.}, \enquote{Broadband huygens’ metasurface based on hybrid resonances,} {\protect\JournalTitle{Physical Review Applied}} \textbf{10}, 034026 (2018).

\bibitem{pancharatnam1956generalized}
S.~Pancharatnam, \enquote{Generalized theory of interference, and its applications: Part i. coherent pencils,} in \emph{Proceedings of the Indian Academy of Sciences-Section A,}  vol.~44 (Springer, 1956), pp. 247--262.

\bibitem{berry1987adiabatic}
M.~V. Berry, \enquote{The adiabatic phase and pancharatnam's phase for polarized light,} {\protect\JournalTitle{Journal of Modern Optics}} \textbf{34}, 1401--1407 (1987).

\bibitem{wang2023metasurface}
S.~Wang, S.~Wen, Z.-L. Deng, \emph{et~al.}, \enquote{Metasurface-based solid poincar{\'e} sphere polarizer,} {\protect\JournalTitle{Physical Review Letters}} \textbf{130}, 123801 (2023).

\bibitem{ren2015generalized}
Z.-C. Ren, L.-J. Kong, S.-M. Li, \emph{et~al.}, \enquote{Generalized poincar{\'e} sphere,} {\protect\JournalTitle{Optics express}} \textbf{23}, 26586--26595 (2015).

\bibitem{yi2015hybrid}
X.~Yi, Y.~Liu, X.~Ling, \emph{et~al.}, \enquote{Hybrid-order poincar{\'e} sphere,} {\protect\JournalTitle{Physical Review A}} \textbf{91}, 023801 (2015).

\bibitem{lin2018polarization}
D.~Lin, A.~L. Holsteen, E.~Maguid, \emph{et~al.}, \enquote{Polarization-independent metasurface lens employing the pancharatnam-berry phase,} {\protect\JournalTitle{Optics Express}} \textbf{26}, 24835--24842 (2018).

\bibitem{ding2015ultrathin}
X.~Ding, F.~Monticone, K.~Zhang, \emph{et~al.}, \enquote{Ultrathin pancharatnam--berry metasurface with maximal cross-polarization efficiency,} {\protect\JournalTitle{Advanced materials}} \textbf{27}, 1195--1200 (2015).

\bibitem{tymchenko2015gradient}
M.~Tymchenko, J.~S. Gomez-Diaz, J.~Lee, \emph{et~al.}, \enquote{Gradient nonlinear pancharatnam-berry metasurfaces,} {\protect\JournalTitle{Physical review letters}} \textbf{115}, 207403 (2015).

\bibitem{xie2021generalized}
X.~Xie, M.~Pu, J.~Jin, \emph{et~al.}, \enquote{Generalized pancharatnam-berry phase in rotationally symmetric meta-atoms,} {\protect\JournalTitle{Physical Review Letters}} \textbf{126}, 183902 (2021).

\bibitem{mcdonnell2021functional}
C.~McDonnell, J.~Deng, S.~Sideris, \emph{et~al.}, \enquote{Functional thz emitters based on pancharatnam-berry phase nonlinear metasurfaces,} {\protect\JournalTitle{Nature communications}} \textbf{12}, 30 (2021).

\bibitem{choudhury2017pancharatnam}
S.~Choudhury, U.~Guler, A.~Shaltout, \emph{et~al.}, \enquote{Pancharatnam--berry phase manipulating metasurface for visible color hologram based on low loss silver thin film,} {\protect\JournalTitle{Advanced Optical Materials}} \textbf{5}, 1700196 (2017).

\bibitem{wang2022nonlinear}
M.~Wang, Y.~Li, Y.~Tang, \emph{et~al.}, \enquote{Nonlinear chiroptical holography with pancharatnam--berry phase controlled plasmonic metasurface,} {\protect\JournalTitle{Laser \& Photonics Reviews}} \textbf{16}, 2200350 (2022).

\bibitem{xu2022mechanically}
Q.~Xu, X.~Su, X.~Zhang, \emph{et~al.}, \enquote{Mechanically reprogrammable pancharatnam--berry metasurface for microwaves,} {\protect\JournalTitle{Advanced Photonics}} \textbf{4}, 016002--016002 (2022).

\bibitem{cai2024compact}
G.~Cai, Y.~Li, Y.~Zhang, \emph{et~al.}, \enquote{Compact angle-resolved metasurface spectrometer,} {\protect\JournalTitle{Nature Materials}} \textbf{23}, 71--78 (2024).

\bibitem{ji2025compact}
W.~Ji, J.~Chang, B.~Mirzaei, \emph{et~al.}, \enquote{Compact metasurface terahertz spectrometer,} {\protect\JournalTitle{Laser \& Photonics Reviews}} \textbf{19}, 2401290 (2025).

\bibitem{wang2023compact}
R.~Wang, M.~A. Ansari, H.~Ahmed, \emph{et~al.}, \enquote{Compact multi-foci metalens spectrometer,} {\protect\JournalTitle{Light: Science \& Applications}} \textbf{12}, 103 (2023).

\bibitem{faraji2018compact}
M.~Faraji-Dana, E.~Arbabi, A.~Arbabi, \emph{et~al.}, \enquote{Compact folded metasurface spectrometer,} {\protect\JournalTitle{Nature communications}} \textbf{9}, 4196 (2018).

\bibitem{tang2024metasurface}
F.~Tang, J.~Wu, T.~Albrow-Owen, \emph{et~al.}, \enquote{Metasurface spectrometers beyond resolution-sensitivity constraints,} {\protect\JournalTitle{Science Advances}} \textbf{10}, eadr7155 (2024).

\bibitem{wen2024metasurface}
S.~Wen, X.~Xue, S.~Wang, \emph{et~al.}, \enquote{Metasurface array for single-shot spectroscopic ellipsometry,} {\protect\JournalTitle{Light: Science \& Applications}} \textbf{13}, 88 (2024).

\bibitem{liu2020optical}
X.~Liu, J.~Deng, K.~F. Li, \emph{et~al.}, \enquote{Optical telescope with cassegrain metasurfaces,} {\protect\JournalTitle{Nanophotonics}} \textbf{9}, 3263--3269 (2020).

\bibitem{zhang2022high}
L.~Zhang, S.~Chang, X.~Chen, \emph{et~al.}, \enquote{High-efficiency, 80 mm aperture metalens telescope,} {\protect\JournalTitle{Nano letters}} \textbf{23}, 51--57 (2022).

\bibitem{wang2025portable}
J.~Wang, Y.~Deng, C.~Wang, \emph{et~al.}, \enquote{Portable astronomical observation system based on large-aperture concentric-ring metalens,} {\protect\JournalTitle{Light: Science \& Applications}} \textbf{14}, 2 (2025).

\bibitem{gopakumar2024full}
M.~Gopakumar, G.-Y. Lee, S.~Choi, \emph{et~al.}, \enquote{Full-colour 3d holographic augmented-reality displays with metasurface waveguides,} {\protect\JournalTitle{Nature}} \textbf{629}, 791--797 (2024).

\bibitem{choi2025roll}
M.~Choi, J.~Kim, S.~Moon, \emph{et~al.}, \enquote{Roll-to-plate printable rgb achromatic metalens for wide-field-of-view holographic near-eye displays,} {\protect\JournalTitle{Nature Materials}} pp. 1--9 (2025).

\bibitem{song2021large}
W.~Song, X.~Liang, S.~Li, \emph{et~al.}, \enquote{Large-scale huygens’ metasurfaces for holographic 3d near-eye displays,} {\protect\JournalTitle{Laser \& Photonics Reviews}} \textbf{15}, 2000538 (2021).

\bibitem{wang2024holographic}
X.~Wang, S.~Liu, L.~Xu, \emph{et~al.}, \enquote{A holographic broadband achromatic metalens,} {\protect\JournalTitle{Laser \& Photonics Reviews}} \textbf{18}, 2300880 (2024).

\bibitem{xiong20253d}
H.~Xiong, X.~Zhang, P.~Li, \emph{et~al.}, \enquote{3d multiview holographic display with wide field of view based on metasurface,} {\protect\JournalTitle{Advanced Optical Materials}} \textbf{13}, 2402504 (2025).

\bibitem{fan2024integral}
Z.-B. Fan, Y.-F. Cheng, Z.-M. Chen, \emph{et~al.}, \enquote{Integral imaging near-eye 3d display using a nanoimprint metalens array,} {\protect\JournalTitle{ELight}} \textbf{4}, 3 (2024).

\bibitem{fan2024dual}
Y.~Fan, H.~Liang, Y.~Wang, \emph{et~al.}, \enquote{Dual-channel quantum meta-hologram for display,} {\protect\JournalTitle{Advanced Photonics Nexus}} \textbf{3}, 016011--016011 (2024).

\bibitem{bouchal2019high}
P.~Bouchal, P.~Dvorak, J.~Babocky, \emph{et~al.}, \enquote{High-resolution quantitative phase imaging of plasmonic metasurfaces with sensitivity down to a single nanoantenna,} {\protect\JournalTitle{Nano letters}} \textbf{19}, 1242--1250 (2019).

\bibitem{wirth2025wide}
A.~Wirth-Singh, J.~E. Fr{\"o}ch, F.~Yang, \emph{et~al.}, \enquote{Wide field of view large aperture meta-doublet eyepiece,} {\protect\JournalTitle{Light: Science \& Applications}} \textbf{14}, 17 (2025).

\bibitem{wang2021metalens}
C.~Wang, Z.~Yu, Q.~Zhang, \emph{et~al.}, \enquote{Metalens eyepiece for 3d holographic near-eye display,} {\protect\JournalTitle{Nanomaterials}} \textbf{11}, 1920 (2021).

\bibitem{li2022ultracompact}
Y.~Li, S.~Chen, H.~Liang, \emph{et~al.}, \enquote{Ultracompact multifunctional metalens visor for augmented reality displays,} {\protect\JournalTitle{PhotoniX}} \textbf{3}, 29 (2022).

\bibitem{lee2018metasurface}
G.-Y. Lee, J.-Y. Hong, S.~Hwang, \emph{et~al.}, \enquote{Metasurface eyepiece for augmented reality,} {\protect\JournalTitle{Nature communications}} \textbf{9}, 4562 (2018).

\bibitem{ko2024metasurface}
J.~Ko, G.~Kim, I.~Kim, \emph{et~al.}, \enquote{Metasurface-embedded contact lenses for holographic light projection,} {\protect\JournalTitle{Advanced Science}} \textbf{11}, 2407045 (2024).

\bibitem{genevet2017recent}
P.~Genevet, F.~Capasso, F.~Aieta, \emph{et~al.}, \enquote{Recent advances in planar optics: from plasmonic to dielectric metasurfaces,} {\protect\JournalTitle{Optica}} \textbf{4}, 139--152 (2017).

\bibitem{qiao2018recent}
P.~Qiao, W.~Yang, and C.~J. Chang-Hasnain, \enquote{Recent advances in high-contrast metastructures, metasurfaces, and photonic crystals,} {\protect\JournalTitle{Advances in Optics and Photonics}} \textbf{10}, 180--245 (2018).

\bibitem{li2017nonlinear}
G.~Li, L.~Wu, K.~F. Li, \emph{et~al.}, \enquote{Nonlinear metasurface for simultaneous control of spin and orbital angular momentum in second harmonic generation,} {\protect\JournalTitle{Nano letters}} \textbf{17}, 7974--7979 (2017).

\bibitem{wang2017metasurface}
C.~Wang, Z.~Li, M.-H. Kim, \emph{et~al.}, \enquote{Metasurface-assisted phase-matching-free second harmonic generation in lithium niobate waveguides,} {\protect\JournalTitle{Nature Communications}} \textbf{8}, 2098 (2017).

\bibitem{ma2021nonlinear}
J.~Ma, F.~Xie, W.~Chen, \emph{et~al.}, \enquote{Nonlinear lithium niobate metasurfaces for second harmonic generation,} {\protect\JournalTitle{Laser \& Photonics Reviews}} \textbf{15}, 2000521 (2021).

\bibitem{liu2018extreme}
M.~Liu and D.-Y. Choi, \enquote{Extreme huygens’ metasurfaces based on quasi-bound states in the continuum,} {\protect\JournalTitle{Nano letters}} \textbf{18}, 8062--8069 (2018).

\bibitem{chen2018phase}
S.~Chen, Z.~Li, Y.~Zhang, \emph{et~al.}, \enquote{Phase manipulation of electromagnetic waves with metasurfaces and its applications in nanophotonics,} {\protect\JournalTitle{Advanced Optical Materials}} \textbf{6}, 1800104 (2018).

\bibitem{yu2011light}
N.~Yu, P.~Genevet, M.~A. Kats, \emph{et~al.}, \enquote{Light propagation with phase discontinuities: generalized laws of reflection and refraction,} {\protect\JournalTitle{science}} \textbf{334}, 333--337 (2011).

\bibitem{so2023revisiting}
S.~So, J.~Mun, J.~Park, and J.~Rho, \enquote{Revisiting the design strategies for metasurfaces: fundamental physics, optimization, and beyond,} {\protect\JournalTitle{Advanced Materials}} \textbf{35}, 2206399 (2023).

\bibitem{chen2012dual}
X.~Chen, L.~Huang, H.~M{\"u}hlenbernd, \emph{et~al.}, \enquote{Dual-polarity plasmonic metalens for visible light,} {\protect\JournalTitle{Nature communications}} \textbf{3}, 1198 (2012).

\bibitem{platt2001history}
B.~C. Platt and R.~Shack, \enquote{History and principles of shack-hartmann wavefront sensing,}  (2001).

\bibitem{yang2024large}
J.~Yang, J.~Zhou, L.~Qiu, \emph{et~al.}, \enquote{Large dynamic range shack-hartmann wavefront sensor based on adaptive spot matching,} {\protect\JournalTitle{Light: Advanced Manufacturing}} \textbf{4}, 42--51 (2024).

\bibitem{he2022high}
F.~He, K.~Liu, Y.~Li, \emph{et~al.}, \enquote{High noise-resistant slope and curvature signal extraction algorithm in frequency domain for the shack-hartmann wavefront sensor,} in \emph{Thirteenth International Conference on Information Optics and Photonics (CIOP 2022),}  vol. 12478 (SPIE, 2022), pp. 870--880.

\bibitem{swanson2018wavefront}
R.~Swanson, M.~Lamb, C.~Correia, \emph{et~al.}, \enquote{Wavefront reconstruction and prediction with convolutional neural networks,} in \emph{Adaptive Optics Systems VI,}  vol. 10703 (SPIE, 2018), pp. 481--490.

\bibitem{li2018centroid}
Z.~Li and X.~Li, \enquote{Centroid computation for shack-hartmann wavefront sensor in extreme situations based on artificial neural networks,} {\protect\JournalTitle{Optics Express}} \textbf{26}, 31675--31692 (2018).

\bibitem{medsker2012hybrid}
L.~R. Medsker, \emph{Hybrid intelligent systems} (Springer Science \& Business Media, 2012).

\bibitem{hu2019learning}
L.~Hu, S.~Hu, W.~Gong, and K.~Si, \enquote{Learning-based shack-hartmann wavefront sensor for high-order aberration detection,} {\protect\JournalTitle{Optics express}} \textbf{27}, 33504--33517 (2019).

\bibitem{dubose2020intensity}
T.~B. DuBose, D.~F. Gardner, and A.~T. Watnik, \enquote{Intensity-enhanced deep network wavefront reconstruction in shack--hartmann sensors,} {\protect\JournalTitle{Optics Letters}} \textbf{45}, 1699--1702 (2020).

\bibitem{guo2022adaptive}
Y.~Guo, L.~Zhong, L.~Min, \emph{et~al.}, \enquote{Adaptive optics based on machine learning: a review,} {\protect\JournalTitle{Opto-Electronic Advances}} \textbf{5}, 200082--1 (2022).

\bibitem{saita2015holographic}
Y.~Saita, H.~Shinto, and T.~Nomura, \enquote{Holographic shack--hartmann wavefront sensor based on the correlation peak displacement detection method for wavefront sensing with large dynamic range,} {\protect\JournalTitle{Optica}} \textbf{2}, 411--415 (2015).

\bibitem{lechner2020adaptable}
D.~Lechner, A.~Zepp, M.~Eichhorn, and S.~G{\l}adysz, \enquote{Adaptable shack-hartmann wavefront sensor with diffractive lenslet arrays to mitigate the effects of scintillation,} {\protect\JournalTitle{Optics express}} \textbf{28}, 36188--36205 (2020).

\bibitem{zhang2024shack}
F.~Zhang, H.~Shen, and Y.~Sun, \enquote{Shack-hartmann wavefront sensor based on a two-dimensional mixed aperture diffractive lens array,} {\protect\JournalTitle{Optics Express}} \textbf{32}, 30724--30741 (2024).

\bibitem{yang2018generalized}
Z.~Yang, Z.~Wang, Y.~Wang, \emph{et~al.}, \enquote{Generalized hartmann-shack array of dielectric metalens sub-arrays for polarimetric beam profiling,} {\protect\JournalTitle{Nature communications}} \textbf{9}, 4607 (2018).

\bibitem{wang2020dielectric}
Y.~Wang, Z.~Wang, X.~Feng, \emph{et~al.}, \enquote{Dielectric metalens-based hartmann--shack array for a high-efficiency optical multiparameter detection system,} {\protect\JournalTitle{Photonics Research}} \textbf{8}, 482--489 (2020).

\bibitem{go2024meta}
G.-H. Go, D.-g. Lee, J.~Oh, \emph{et~al.}, \enquote{Meta shack--hartmann wavefront sensor with large sampling density and large angular field of view: phase imaging of complex objects,} {\protect\JournalTitle{Light: Science \& Applications}} \textbf{13}, 187 (2024).

\bibitem{torti2008wavefront}
C.~Torti, S.~Gruppetta, and L.~Diaz-Santana, \enquote{Wavefront curvature sensing for the human eye,} {\protect\JournalTitle{Journal of Modern Optics}} \textbf{55}, 691--702 (2008).

\bibitem{xin2015curvature}
B.~Xin, C.~Claver, M.~Liang, \emph{et~al.}, \enquote{Curvature wavefront sensing for the large synoptic survey telescope,} {\protect\JournalTitle{Applied Optics}} \textbf{54}, 9045--9054 (2015).

\bibitem{guyon2009high}
O.~Guyon, \enquote{High sensitivity wavefront sensing with a nonlinear curvature wavefront sensor,} {\protect\JournalTitle{Publications of the Astronomical Society of the Pacific}} \textbf{122}, 49 (2009).

\bibitem{ahn2023development}
K.~Ahn, O.~Guyon, J.~Lozi, \emph{et~al.}, \enquote{Development of a non-linear curvature wavefront sensor for the subaru telescope's ao3k system,} in \emph{Adaptive Optics for Extremely Large Telescopes 7th Edition,}  (2023).

\bibitem{jimenez2024single}
A.~M. Jimenez, M.~Baltes, J.~Cornelius, \emph{et~al.}, \enquote{Single-shot phase diversity wavefront sensing in deep turbulence via metasurface optics,} {\protect\JournalTitle{arXiv preprint arXiv:2410.18789}}  (2024).

\bibitem{zhou2022single}
H.~Zhou, X.~Li, N.~Ullah, \emph{et~al.}, \enquote{Single-shot phase retrieval based on anisotropic metasurface,} {\protect\JournalTitle{Applied Physics Letters}} \textbf{120} (2022).

\bibitem{zhou2025advanced}
J.~Zhou, A.~Li, M.~Lei, \emph{et~al.}, \enquote{Advanced quantitative phase microscopy achieved with spatial multiplexing and a metasurface,} {\protect\JournalTitle{Nano Letters}}  (2025).

\bibitem{liu2024metalenses}
B.~Liu, J.~Cheng, M.~Zhao, \emph{et~al.}, \enquote{Metalenses phase characterization by multi-distance phase retrieval,} {\protect\JournalTitle{Light: Science \& Applications}} \textbf{13}, 182 (2024).

\bibitem{cheng2025quantitative}
J.~Cheng, Y.~Zhou, Y.~Gao, \emph{et~al.}, \enquote{Quantitative phase imaging for meta-lenses by phase retrieval,} {\protect\JournalTitle{Advanced Optical Materials}} \textbf{13}, 2402833 (2025).

\bibitem{primot2000extended}
J.~Primot and N.~Gu{\'e}rineau, \enquote{Extended hartmann test based on the pseudoguiding property of a hartmann mask completed by a phase chessboard,} {\protect\JournalTitle{Applied optics}} \textbf{39}, 5715--5720 (2000).

\bibitem{chanteloup2005multiple}
J.-C. Chanteloup, \enquote{Multiple-wave lateral shearing interferometry for wave-front sensing,} {\protect\JournalTitle{Applied optics}} \textbf{44}, 1559--1571 (2005).

\bibitem{karp2008integrated}
J.~H. Karp, T.~K. Chan, and J.~E. Ford, \enquote{Integrated diffractive shearing interferometry for adaptive wavefront sensing,} {\protect\JournalTitle{Applied Optics}} \textbf{47}, 6666--6674 (2008).

\bibitem{ling2015general}
T.~Ling, Y.~Yang, D.~Liu, \emph{et~al.}, \enquote{General measurement of optical system aberrations with a continuously variable lateral shear ratio by a randomly encoded hybrid grating,} {\protect\JournalTitle{Applied Optics}} \textbf{54}, 8913--8920 (2015).

\bibitem{ling2015quadriwave}
T.~Ling, D.~Liu, X.~Yue, \emph{et~al.}, \enquote{Quadriwave lateral shearing interferometer based on a randomly encoded hybrid grating,} {\protect\JournalTitle{Optics Letters}} \textbf{40}, 2245--2248 (2015).

\bibitem{zhong2024phase}
H.~Zhong, Y.~Li, K.~Liu, \emph{et~al.}, \enquote{Phase microscopy using the quadriwave lateral shearing interferometer with continuously variable shear ratio,} in \emph{AOPC 2024: Computational Imaging Technology,}  vol. 13501 (SPIE, 2024), pp. 87--92.

\bibitem{zhang2019novel}
R.~Zhang, Y.~Yang, and Z.~Liang, \enquote{A novel aspherical surface measurement system based on a randomly encoded hybrid grating wavefront sensor,} in \emph{Applied Optical Metrology III,}  vol. 11102 (SPIE, 2019), pp. 48--54.

\bibitem{agour2022characterization}
M.~Agour, C.~Falldorf, F.~Taleb, \emph{et~al.}, \enquote{Characterization of terahertz wavefront aberrations using computational shear-interferometry,} {\protect\JournalTitle{Optical Engineering}} \textbf{61}, 114102--114102 (2022).

\bibitem{jiang2016measurement}
J.~Jiang, T.~Ling, Y.~Yang, and R.~Zhang, \enquote{Measurement of optical system aberrations based on randomly encoded hybrid grating,} in \emph{Optical Design and Testing VII,}  vol. 10021 (SPIE, 2016), pp. 215--221.

\bibitem{munj2023unidirectional}
A.-H. Munj and W.~R. Kerridge-Johns, \enquote{Unidirectional ring vortex laser using a wedge-plate shearing interferometer,} {\protect\JournalTitle{Optics Express}} \textbf{31}, 4954--4963 (2023).

\bibitem{guo2020quantitative}
R.~Guo, S.~K. Mirsky, I.~Barnea, \emph{et~al.}, \enquote{Quantitative phase imaging by wide-field interferometry with variable shearing distance uncoupled from the off-axis angle,} {\protect\JournalTitle{Optics express}} \textbf{28}, 5617--5628 (2020).

\bibitem{zhang2019ultrafast}
S.~Zhang, P.~Li, S.~Wang, \emph{et~al.}, \enquote{Ultrafast vortices generation at low pump power and shearing interferometer-based vortex topological detection,} {\protect\JournalTitle{Laser Physics Letters}} \textbf{16}, 035302 (2019).

\bibitem{agour2022terahertz}
M.~Agour, C.~Fallorf, F.~Taleb, \emph{et~al.}, \enquote{Terahertz referenceless wavefront sensing by means of computational shear-interferometry,} {\protect\JournalTitle{Optics Express}} \textbf{30}, 7068--7081 (2022).

\bibitem{liu2018high}
Y.~Liu, M.~Seaberg, D.~Zhu, \emph{et~al.}, \enquote{High-accuracy wavefront sensing for x-ray free electron lasers,} {\protect\JournalTitle{Optica}} \textbf{5}, 967--975 (2018).

\bibitem{nagler2017focal}
B.~Nagler, A.~Aquila, S.~Boutet, \emph{et~al.}, \enquote{Focal spot and wavefront sensing of an x-ray free electron laser using ronchi shearing interferometry,} {\protect\JournalTitle{Scientific reports}} \textbf{7}, 13698 (2017).

\bibitem{makita2020double}
M.~Makita, G.~Seniutinas, M.~H. Seaberg, \emph{et~al.}, \enquote{Double grating shearing interferometry for x-ray free-electron laser beams,} {\protect\JournalTitle{Optica}} \textbf{7}, 404--409 (2020).

\bibitem{goldberg2020reflective}
K.~A. Goldberg, D.~Bryant, A.~Wojdyla, \emph{et~al.}, \enquote{Reflective binary amplitude grating for soft x-ray shearing and hartmann wavefront sensing,} {\protect\JournalTitle{Optics Letters}} \textbf{45}, 4694--4697 (2020).

\bibitem{goldberg2021binary}
K.~A. Goldberg, A.~Wojdyla, and D.~Bryant, \enquote{Binary amplitude reflection gratings for x-ray shearing and hartmann wavefront sensors,} {\protect\JournalTitle{Sensors}} \textbf{21}, 536 (2021).

\bibitem{song2024extended}
I.-U. Song, J.-H. Choi, H.-G. Rhee, \emph{et~al.}, \enquote{Extended wavefront reconstruction for quadri-wave lateral shearing interferometry,} {\protect\JournalTitle{Optics and Lasers in Engineering}} \textbf{178}, 108212 (2024).

\bibitem{peng2020calibration}
C.~Peng, F.~Tang, X.~Wang, and J.~Li, \enquote{Calibration method of shear amount based on the optical layout of point source microscope for lateral shearing interferometric wavefront sensor,} {\protect\JournalTitle{Optical Engineering}} \textbf{59}, 094106--094106 (2020).

\bibitem{lam2018complete}
B.~Lam and C.~Guo, \enquote{Complete characterization of ultrashort optical pulses with a phase-shifting wedged reversal shearing interferometer,} {\protect\JournalTitle{Light: Science \& Applications}} \textbf{7}, 30 (2018).

\bibitem{song2022surface}
I.-U. Song, H.-S. Yang, G.~Kim, and S.-W. Kim, \enquote{Surface form error measurement for rough surfaces using an infrared lateral shearing interferometry,} {\protect\JournalTitle{Optics and Lasers in Engineering}} \textbf{152}, 106947 (2022).

\bibitem{wu2023single}
Q.~Wu, J.~Zhou, X.~Chen, \emph{et~al.}, \enquote{Single-shot quantitative amplitude and phase imaging based on a pair of all-dielectric metasurfaces,} {\protect\JournalTitle{Optica}} \textbf{10}, 619--625 (2023).

\bibitem{li2024single}
L.~Li, S.~Wang, F.~Zhao, \emph{et~al.}, \enquote{Single-shot deterministic complex amplitude imaging with a single-layer metalens,} {\protect\JournalTitle{Science Advances}} \textbf{10}, eadl0501 (2024).

\bibitem{endrizzi2018x}
M.~Endrizzi, \enquote{X-ray phase-contrast imaging,} {\protect\JournalTitle{Nuclear instruments and methods in physics research section A: Accelerators, spectrometers, detectors and associated equipment}} \textbf{878}, 88--98 (2018).

\bibitem{kim2022spiral}
Y.~Kim, G.-Y. Lee, J.~Sung, \emph{et~al.}, \enquote{Spiral metalens for phase contrast imaging,} {\protect\JournalTitle{Advanced Functional Materials}} \textbf{32}, 2106050 (2022).

\bibitem{xingmonolithic}
Z.~Xing, Z.~Lin, N.~Liu, \emph{et~al.}, \enquote{Monolithic spin-multiplexing metalens for dual-functional imaging,} {\protect\JournalTitle{Laser \& Photonics Reviews}} p. 2401993 (2025).

\bibitem{ji2022quantitative}
A.~Ji, J.-H. Song, Q.~Li, \emph{et~al.}, \enquote{Quantitative phase contrast imaging with a nonlocal angle-selective metasurface,} {\protect\JournalTitle{Nature Communications}} \textbf{13}, 7848 (2022).

\bibitem{fu2019measuring}
Y.~Fu, C.~Min, J.~Yu, \emph{et~al.}, \enquote{Measuring phase and polarization singularities of light using spin-multiplexing metasurfaces,} {\protect\JournalTitle{Nanoscale}} \textbf{11}, 18303--18310 (2019).

\bibitem{pelzman2019wavefront}
C.~Pelzman and S.-Y. Cho, \enquote{Wavefront detection using curved nanoscale apertures,} {\protect\JournalTitle{Applied Physics Letters}} \textbf{114} (2019).

\bibitem{aharonov1988result}
Y.~Aharonov, D.~Z. Albert, and L.~Vaidman, \enquote{How the result of a measurement of a component of the spin of a spin-1/2 particle can turn out to be 100,} {\protect\JournalTitle{Physical review letters}} \textbf{60}, 1351 (1988).

\bibitem{luo2024phase}
W.~Luo, Q.~Yang, Y.~Wang, \emph{et~al.}, \enquote{Phase and amplitude reconstruction by weak measurement based on metasurface,} {\protect\JournalTitle{Applied Physics Letters}} \textbf{124} (2024).

\bibitem{ballew2023multi}
C.~Ballew, G.~Roberts, and A.~Faraon, \enquote{Multi-dimensional wavefront sensing using volumetric meta-optics,} {\protect\JournalTitle{Optics Express}} \textbf{31}, 28658--28669 (2023).

\bibitem{jensen2011topology}
J.~S. Jensen and O.~Sigmund, \enquote{Topology optimization for nano-photonics,} {\protect\JournalTitle{Laser \& Photonics Reviews}} \textbf{5}, 308--321 (2011).

\bibitem{ji2023recent}
W.~Ji, J.~Chang, H.-X. Xu, \emph{et~al.}, \enquote{Recent advances in metasurface design and quantum optics applications with machine learning, physics-informed neural networks, and topology optimization methods,} {\protect\JournalTitle{Light: Science \& Applications}} \textbf{12}, 169 (2023).

\bibitem{dainese2024shape}
P.~Dainese, L.~Marra, D.~Cassara, \emph{et~al.}, \enquote{Shape optimization for high efficiency metasurfaces: theory and implementation,} {\protect\JournalTitle{Light: Science \& Applications}} \textbf{13}, 300 (2024).

\bibitem{phan2019high}
T.~Phan, D.~Sell, E.~W. Wang, \emph{et~al.}, \enquote{High-efficiency, large-area, topology-optimized metasurfaces,} {\protect\JournalTitle{Light: Science \& Applications}} \textbf{8}, 48 (2019).

\end{thebibliography}






\end{document}